\documentclass[twocolumn,aps,prb,superscriptaddress]{revtex4-1}
\usepackage{amsmath,amssymb}
\usepackage{graphicx,xcolor,soul}
\usepackage[colorlinks,bookmarks=false,citecolor=blue,linkcolor=blue,urlcolor=blue,hyperfootnotes=true]{hyperref}
\makeatother

\usepackage{lineno}

\begin{document}
\title{Spin current injection via equal-spin Cooper pairs in ferromagnet/superconductor heterostructures}

\author{X. Montiel}
\email[]{xm252@cam.ac.uk}
\affiliation{Department of Materials Science \& Metallurgy, University of Cambridge, CB3 0FS Cambridge, United Kingdom}
\author{M. Eschrig} 
\email[]{matthias.eschrig@uni-greifswald.de}
\affiliation{Institute of Physics, University of Greifswald, Felix-Hausdorff-Strasse 6, 17489 Greifswald, Germany}

\date{\today}
\begin{abstract}
Equal-spin Cooper pairs are pivotal building blocks for superconducting spintronics devices. In recent experiments unusual behavior was observed in ferromagnet/ferromagnet/superconductor devices when a precession of the magnetization was induced by ferromagnetic resonance. 
By using a non-equilibrium Usadel Green function formalism, we study spin transport for such a setup. We solve for spin-resolved distribution functions and demonstrate that the spin injection process in superconductors is governed by the inverse proximity effect in the superconducting layer. We find that equal-spin Cooper pairs, which are produced by the two misaligned ferromagnetic layers, transport spin inside the S layer. This then results in an increase of the injected spin current below the superconducting critical temperature. Our calculations provide the first evidence of the essential role of equal-spin Cooper pairs on spin-transport properties of S/F devices and pave new avenues for the design of superconducting spintronics devices.
\end{abstract}
\maketitle
\noindent

\textit{Introduction}
Spin transport in superconductors (S) is essential for the development of energy-saving spin-based logic and memory devices \cite{Eschrig_PhysTod2011,Eschrig_RepProgPhys2015,Robinson_Linder_2015}. Contrary to conventional singlet Cooper pairs, composed by two electrons with opposite spins, equal-spin cooper pairs are composed by two electrons with parallel spins, and are able to carry a spin polarized current through strongly spin polarized ferromagnets (F) \cite{Eschrig_PhysTod2011,Eschrig_RepProgPhys2015}. Equal-spin Cooper pairs can be produced from singlet Cooper pairs by proximity effect with inhomogeneous magnetic structures \cite{Izyumov2002,Eschrig_adv2004,Golubov_RevMod2004,Bergeret_RevMod2005,Buzdin_RevMod2005,Lyuksyutov2005,Eschrig_PhysTod2011,BlamireRobinson_JPhys2014,Eschrig_RepProgPhys2015,Robinson_Linder_2015,Birge_PhilTransA2018,Sauls_PRB1988,Fogelstrom00,BobkovaBobkov02,Eschrig_PRL2003,Bergeret_PRL2001} (e.g. two ferromagnets with non colinear magnetizations \cite{Champel_PRL2005,houzet_PRB2007,Halterman_PRB2008} or magnetic domains wall \cite{Champel_PRB2005,Champel_PRB2005b,Fominov_PRB2007,Crouzy_PRB2007,Pugach_PRB2011,Kupferschmidt_PRB2009,Bergeret_PRL2001}), strongly spin polarized ferromagnet with spin-active interfaces \cite{Eschrig_PRL2003,eschrig_NatPhys2008} or material exhibiting spin-orbit couplingv (SOC) \cite{Annunziata_PRB2012,Bergeret_PRL2013,Bergeret_PRB2014,Linder_PRB2015,Zutic_PRL2015}. Non-equilibrium spin injection techniques can be used together with measurements of transport properties to characterize the presence of equal-spin Cooper pairs in mesoscopic devices \cite{Grein_PRB2010,Hubler_PRB2010,Hubler_PRL2012,Hubler_PRB2013,quay_NatPhys2013,Wakamura_PRL2014,Beckmann_JPhysConMat2016}. They affect spin and charge decoupling in superconductors \cite{quay_NatPhys2013,Wakamura_PRL2014}, spin relaxation time \cite{Poli_PRL2008,yang_NatMat2010,wakamura_NatMat2015} or effective spin-orbit interaction \cite{Inoue_PRB2017} but their effect on spin current has not been ubiquitously established.

Spin current injection in a superconductor can be achieved either by the injection of a spin-polarized charge current in lateral structures \cite{quay_NatPhys2013} or by injecting pure spin current utilizing ferromagnetic resonance (FMR) technique \cite{Bell_PRL2008,Jeon_NatMat2018}. In the latter, spin injection efficiency is related to the damping of the F layer magnetization precession i.e. the Gilbert damping \cite{Tserkovnyak_RevModPhys2005}. FMR experiments in Nb/Py/Nb Josephson junctions have shown that the injected spin current magnitude (related to Gilbert damping \cite{Tserkovnyak_RevModPhys2005}) decreases below the S layer critical temperature $T_c$ \cite{Bell_PRL2008}. Usual Andreev reflections occurring at the S/F interfaces imply the transmission of electrons in the S layer as singlet Cooper pairs resulting in a decrease of the injected spin-polarized current \cite{Morten_EPL2008}. 
Nevertheless, FMR experiments in Pt/Nb/Py/Nb/Pt pentalayers have shown an increase of the injected spin current below $T_c$ \cite{Jeon_NatMat2018}.
This increase has been explained by the presence of equal-spin Cooper pairs originating the strong SOC and Fermi liquid corrections in the Pt layer \cite{Montiel_PRB2018}.
Several extrinsic sources of damping have been proposed such as spin-polarized vortices \cite{Tserkovnyak_PRL_2018} or space-dependent spin susceptibility \cite{Taira_PRB_2018}. Recent experiments in Pt/Co/Pt/Nb/Py/Nb/Pt/Co/Pt structure have demonstrated the crucial role of SOC in the increase of spin injection efficiency below $T_c$ \cite{Jeon_PRX_2020}. An increase in the Gilbert damping has been proven as possible in weakly coupled ferromagnetic insulator/superconductor bilayers \cite{Inoue_PRB2017} or in presence of pair breaking impurities \cite{Morten_EPL2008}. In the above-mentioned works, the increase of spin current only exists for temperatures close to $T_c$. While the role of equal-spin Cooper pairs on equilibrium spin current has been demonstrated \cite{Montiel_PRB2018,Jeon_PRX_2020}, their precise role on the non-equilibrium pure spin current is still lacking.  

\begin{figure}
    \centering
    \includegraphics[width=7.5cm]{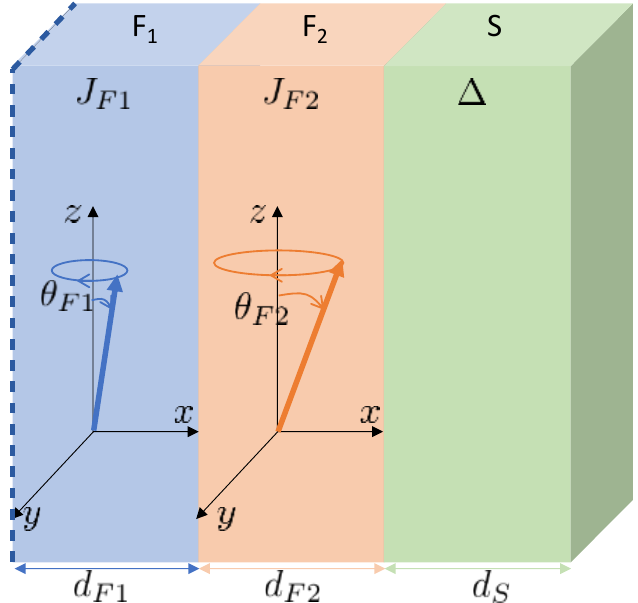}
    \caption{(Color online) Schematic geometry of the F$_1$/F$_2$/S spin valve of thickness $d_{F1}$, $d_{F2}$ and $d_S$ respectively and $L=d_{F1}+d_S+d_{F2}$ .We assume that that both F layers magnetization precess at the same frequency $\Omega$.}
    \label{fig:fig_1}
\end{figure}

In this paper, we calculate non-equilibrium spin current in F1/F2/S devices (see Fig. \ref{fig:fig_1}) induced by the magnetization precession in the presence of equal-spin Cooper pairs. In contrast to modelling the injection process  for a fixed value of the injected spin current \cite{Silaev_PRB2020,Simensen_PRB2021}, we will take another avenue by directly modelling the precessing ferromagnetic magnetization within a rotating frame scheme. Our calculations provide evidence that i) the inverse proximity effect in the S layer plays a crucial role in the spin injection process, and ii) the injected spin current can increase below $T_c$ in presence of equal-spin Cooper pairs. {\color{black} Our model is consistent with the recent observation of an increase of the Gilbert damping at low temperature observed in Nb/Cr/Fe/Cr/Nb pentalayer \cite{Chan_arxiv2022}.}

\textit{Model}
We focus on  spin-valve systems with a geometry shown in Fig.\ref{fig:fig_1} where magnetization precession is induced {\color{black} in both F1 and F2 layers at the same frequency $\Omega$}. We consider spin precession occurring in the x-y plane while magnetization is tilted from the z-axis by the angle $\theta_{F1}$  {\color{black} $(\theta_{F2})$ in F1 (F2) layer. While having the magnetization in the F2 layer non-precessing would certainly be closer to recent experiments,
 the model we study here has the advantage of allowing for a numerically efficient treatment of the spatio-temporal spin dynamics by utilizing a rotating frame scheme}. We model the time-dependence of the ferromagnetic layers magnetization by \cite{Houzet_PRL101_2008} 
 \begin{equation}
\mathbf{J}_{Fi}\left(t\right)=|J_{Fi}|\left(\text{sin}\theta_{Fi}\text{sin}\Omega t ,\text{sin}\theta_{Fi}\text{cos}\Omega t,\text{cos}\theta_{Fi}\right)
    \label{J_dep_t}
\end{equation}
where  {\color{black} $i=1,2$} and $J_{Fi}$ is the Fi layer exchange-field strength. To obtain non-equilibrium and non-stationary properties in diffusive superconductors, we use time-dependent non-equilibrium Usadel equations \cite{Usadel_PRL1970,Belzig_SuperLattice1999,Houzet_PRL101_2008,Eschrig_PRB2009} which are derived from quasi-classical equations \cite{Eilenberger_1968,larkin_JETP1969}. The exchange field time dependency in  Eq.(\ref{J_dep_t}) allows to transform non-stationary Usadel equations in the laboratory frame into stationary Usadel equations in the rotating frame \cite{Houzet_PRL101_2008}. 
Usadel formalism is formulated in terms of Green function $\check{G}\left(\boldsymbol{x},E\right)$ depending on the coordinate $\boldsymbol{x}$ and the energy $E$.
The Green function $\check{G}$ is a $8\times 8$ matrix in the Keldysh$\otimes$spin$\otimes$particle-hole space where $\otimes$ is the tensorial product and Keldysh, spin and particle-hole are $2\times 2$ subspaces (see Appendix \ref{appA}).
The stationary non-equilibrium Usadel equations in the rotating frame are (see Appendix \ref{appB}) \cite{Houzet_PRL101_2008}
\begin{equation}
\frac{D}{\pi}\partial_{x}\left(\check{G}\partial_{x}\check{G}\right)+\left[E\hat{\tau}_{3}-\frac{\Omega}{2}\sigma^{Z}-\check{\Sigma},\check{G}\right]=0
\label{eq_us}
\end{equation}
where $D$ is the diffusion coefficient, $\sigma^Z (\tau_3)$ the third Pauli matrix acting on the spin (particle-hole) subspace, $[,]$ is the commutator and $\check{\Sigma}$ describes the normal and anomalous self-energies in the Keldysh space (see Appendix \ref{appA}). 
{\color{black} The normal self-energy includes spin-orbit scattering processes (see Appendix \ref{appA}). Spin-orbit scattering processes describe how spin scatters on non-magnetic impurities due to spin-orbit coupling \cite{Rusinov_JETP1964,Inoue_PRB2017}.  The strength of these processes increases with the atomic number of the scattering impurity \cite{Rusinov_JETP1964,Inoue_PRB2017}. Spin-orbit scattering processes  exist at all temperatures and do not affect the SC critical temperature as no time-reversal symmetry breaking is involved in these processes.  The superconducting self-energy describes the superconducting order parameter (see Appendix \ref{appA}). Transformation into a rotating frame induces the onset of an effective exchange field along the F1 layer magnetization whose amplitude is proportional to the magnetic field precession frequency $\Omega$.}
We rotate the axis such that the F1 layer magnetization points in the z-axis direction
\cite{Houzet_PRL101_2008} {\color{black} implying all results depend on the difference between the F1 and the F2 layer magnetization precession angle, $\theta=\theta_{F2}-\theta_{F1}$.} Identity matrices are not explicitely written for simplicity of notation.
The Usadel equations (\ref{eq_us}) must be supplemented by suitable boundary conditions \cite{Kuprianov_JETP1988,Nazarov_PRL1994,Nazarov_SuperLattMicro1999,Eschrig_NJPhys2015}. For inner interfaces, we assume current conserving Nazarov boundary conditions \cite{Nazarov_PRL1994,Nazarov_SuperLattMicro1999}:
\begin{equation}
\begin{array}{c}
\sigma_l \check{G}_l \partial_x \check{G}_l=\sigma_r \check{G}_r \partial_x \check{G}_r \\
\sigma_l\check{G}_l \partial_x \check{G}_l=\frac{1}{S R_b}\frac{2\pi^2 \tau \left[\check{G}_l,\check{G}_r\right]}{4\pi^2-\tau \left(\{ \check{G}_l,\check{G}_r\}+2\pi^2\right)}
\end{array}
\label{BC_Naz}
\end{equation}
where $G^{l(r)}$ is the Green function on the left(right) side of the interface, $\sigma$ is the normal state electrical conductivity, S is the area of the junction, $R_b$ is the interface resistivity and $0<\tau<1$ the interface transparency \cite{Nazarov_SuperLattMicro1999,Eschrig_NJPhys2015}. For the outer F1 boundary at $x=0$, we impose $\check{G}(x=0)=\check{G}^{F1}$ where $\check{G}^{F1}$ is the Green function for a bulk ferromagnetic material with a spin-resolved non-equilibrium distribution function $f_{\uparrow(\downarrow)}=f_{FD}(E+(-)\frac{\Omega'}{2})$ with $f_{FD}$ the Fermi Dirac distribution.
At the outer S boundary ($x=L$), we impose a vanishing-current boundary condition $\partial_x  \left. \check{G} \right|_{x=L}=0$. The singlet-superconducting order parameter is fixed by the self-consistency equation
\begin{equation}
\Delta\left(x\right)=\frac{\int_{-\infty}^{\infty}\frac{dE}{4i\pi}f_{s}^{K}\left(E,x\right)}{\int_{-\infty}^{\infty}\frac{dE}{2E}\text{tanh}\left(\frac{E}{2T}\right)+\text{ln}\left(\frac{T}{Tc}\right)}\label{eq:SC_self}
\end{equation}
where $f_s^K$ is the singlet part of the Keldysh anomalous Green function (see Appendix \ref{appA}). In the following, we calculate spin current densities from the following formula:
\begin{equation}
\boldsymbol{I}_s=I_s^0\int_{-\infty}^{+\infty}dE\text{Tr}\left[\hat{\tau_3}\boldsymbol{\sigma}\left(\check{G}\partial_x\check{G}\right)^K\right]
\label{Eq_spincurrent}
\end{equation}
with $I_s^0=\frac{\hbar N_0 D}{16 \pi^2 L}$, $N_0$ the Fermi level density of state, $e$ the electrical charge, $\hbar$ the reduced Planck constant, $L$ the sample total thickness, and $\boldsymbol{\sigma}=\left(\sigma^X,\sigma^Y,\sigma^Z\right)$ the Pauli matrix vector. The spin current vector $\boldsymbol{I}_s$ is given by $\boldsymbol{I}_s=\left(I_s^X,I_s^Y,I_s^Z\right)$ in the Pauli matrix basis. A spin-dependent distribution function $f_{\uparrow(\downarrow)}$ at the outer F1 boundary implies the onset of a pure spin current to flow in the trilayer while the charge current vanishes  since no electric potential is applied \cite{quay_NatPhys2013}. The pure spin current is polarized along the F1 layer magnetization $I_s^Z$.  In absence of magnetization precession ($\Omega=0$), we recover the equilibrium solution and $I_{s,eq}^Z=0$. 

 The Gilbert damping $G_t$ can be expressed in the form
\begin{align}
G_t=G_0+G\equiv G_0+\beta I_s^Z
\label{Gt}
\end{align}
where $G_0$ describes an intrinsic Gilbert damping independent of temperature and superconducting properties, and $G$ describes the additional Gilbert damping due to spin injection; the latter is proportional to the dissipative part of the spin current \cite{Simensen_PRB2021}, which in itself is proportional to $I_s^Z$ at $x=0$ with a coefficient depending on the tip angle. The quantity $\delta G/G_{T>T_c}=\delta I_s^Z/I_{s,T>T_c}^Z$ with $\delta I_s^Z=I_s^Z-I_{s,T>T_c}^Z$ is independent of the tip-angle dependent quantity $\beta $ and therefore quantifies the additional Gilbert damping in our system.

We assume that spin diffusion originates from spin-orbit scattering processes which only affect spin-triplet correlations without affecting spin-singlet superconductivity \cite{Inoue_PRB2017,Morten_EPL2008,Rusinov_JETP1964,Jacobsen_SciRep2016,Jacobsen_PRB2017,Montiel_PRB2018}. We further assume that the FMR process in the F1 layer compensates spin diffusion processes i.e. the spin diffusion length in the F1 layer is infinite, $\lambda_{F1}\rightarrow \infty$. Unless otherwise stated,  the magnitude of the F1 and F2 layer exchange fields is $J_{F1}=J_{F2}=20\Delta_0$, the spin-diffusion length of the F2 and S layer is $\lambda_{F2}=\lambda_{S}=\xi_0$ with $\xi_0=\sqrt{\frac{D}{\Delta_0}}$ the superconducting coherence length in bulk S at zero temperature, and we consider $\xi_0=30\text{nm}$ in Niobium. In the following, we set the resonance frequency at the experimentally measured value $f_{res}=20\text{GHz}$ \cite{Bell_PRL2008,Jeon_NatMat2018} implying that $\hbar\Omega \approx 0.1 \Delta_0$.

\textit{Results}
\begin{figure}
    \centering
    \includegraphics[width=8.6cm]{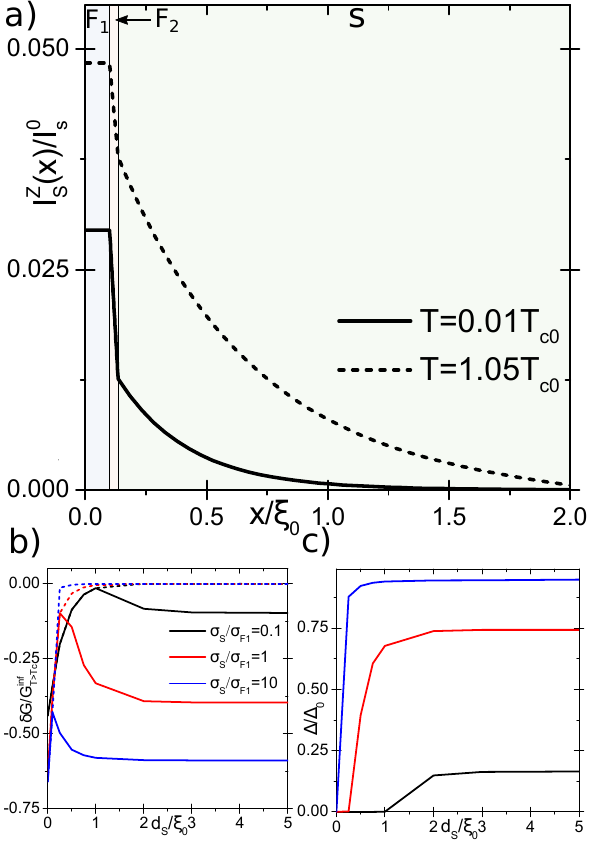}
    \caption{(Color online) a) Spin current profile $I_s^Z(x)$ in the F1/F2/S trilayer (in blue, orange and green respectively) at low temperature $T=0.01T_{c0}$ (solid line) and in the normal state $T>T_{c0}$ (dashed line) for $\theta_{F2}=0$.  
    The layer thicknesses are $d_{F1}=\xi_0$, $d_{F2}=0.1\xi_0$, and $d_{S}=2\xi_0$ while $SR_b=1$ and $\tau=1$. The vertical lines mark the position of the F1/F2 and F2/S interfaces. b) 
    Gilbert damping parameter $\delta G/G_{T>T_c}$ at the outer F1 layer boundary as a function of the S layer thickness $d_S$ for various S layer conductivities (see color legend) below and above $T_c$ (solid and dashed line respectively) c) Magnitude of the superconducting gap at the F2/S interface as a function of the S layer thickness $d_S$ for various S layer conductivities.}
    \label{fig:fig_2}
\end{figure}
The spin current profile in the F1/F2/S trilayer for collinear magnetizations ($\theta=0$) is presented in Fig. \ref{fig:fig_2}. 
The magnitude of the spin current decays in the F2 and S layers because of spin-orbit scattering processes. The magnitude of the spin current is higher in the normal state than in the superconducting state because of the opening of the superconducting gap.
The spin current is constant at the inner interfaces as expected from boundary conditions (\ref{BC_Naz}). The magnitude of the spin current strongly depends on the inverse proximity effect and the S layer thickness $d_S$, which affects the Gilbert damping parameter $\delta G/G_{T>T_c}$ as shown in Fig. \ref{fig:fig_2} b). 
For small S layer thicknesses, $d_S < \xi_0$, the  
Gilbert damping parameter is the same above and below $T_c$ since in both cases the superconducting gap vanishes at the F2/S interfaces (except in the regime $\sigma_S/\sigma_{F1}=10$) as shown in Fig. \ref{fig:fig_2} c). Note that for $d_S=0$, the 
Gilbert damping parameter does not vanish since spins are absorbed in the F2 layer. For a thick S layer, $d_S \gg \xi_0=\lambda_S$, the
Gilbert damping parameter above and below $T_c$ becomes constant since the spin is massively absorbed in the S layer close to the F2/S interface on the spin diffusion length scale $\lambda_S$. 
In the intermediate regime, $d_S \approx \xi_0$, the Gilbert damping parameter below $T_c$ becomes sightly different from above $T_c$ in conjunction with the superconducting gap opening at the F2/S interfaces as shown in Fig. \ref{fig:fig_2} c). The inverse proximity effect can be tuned theoretically by changing the normal state conductivity ratio $\sigma_S/\sigma_{F1}$ (with $\sigma_{F2}=\sigma_{F1}$) \cite{Montiel_PRB2018}.
For a weak inverse proximity effect i.e. $\sigma_S/\sigma_{F1}=10$, the superconducting gap $\Delta$ is fully established at the F2/S interfaces. In the regime $\hbar \Omega<\Delta$, spin currents cannot find any states to propagate further in the S layer. Only singlet Andreev reflections occur at the F2/S interface implying a decay of the Gilbert damping below $T_c$ \cite{Bell_PRL2008,Morten_EPL2008}. For a strong inverse proximity effect i.e. $\sigma_S/\sigma_{F1}=0.1$, the superconducting gap is strongly suppressed at the F2/S interface implying that non-equilibrium spin current can be injected in the S layer. Therefore, the Gilbert damping recovers the same magnitude above and below $T_c$. 
The dependence of the Gilbert damping on the inverse proximity effect explains why it does not necessarily vanishes at zero temperature \cite{Bell_PRL2008,Morten_EPL2008}.
\begin{figure}
    \centering
    \includegraphics[width=8.6cm]{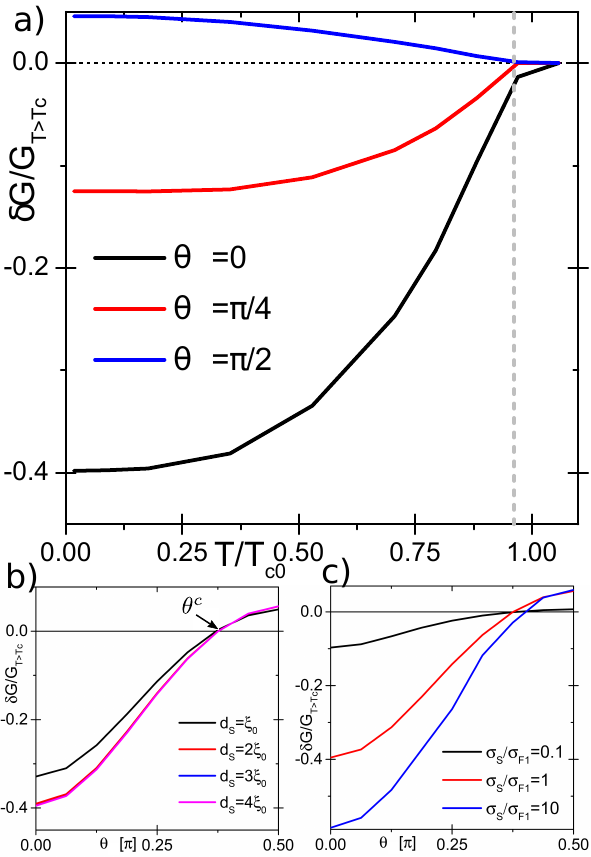}
    \caption{(Color online) a) Temperature dependence of the Gilbert damping parameter $\delta G/G_{T>T_c}$ for $T=0.01T_c$
    for various F2 layer magnetization tilting angles $\theta=0$ (black), $\theta=\pi/4$ (red),and $\theta=\pi/2$ (blue) and $d_S=4\xi_0$. b) Gilbert damping parameter $\delta G/G_{T>T_c}$ as a function of $\theta_{F2}$ for different S layer thicknesses $d_S$. $\theta^c$ is the F2 layer magnetization angle where $\Delta I_s^Z$ changes its sign. c) Gilbert damping parameter $\delta G/G_{T>T_c}$ as a function of $\theta_{F2}$ for various S layer conductivities $\sigma_S$ and $d_S=4\xi_0$. The other parameters are the same as in Fig. \ref{fig:fig_2}}
    \label{fig:fig_3}
\end{figure}

The Gilbert damping varies with temperature as shown in Fig. \ref{fig:fig_3} a). This dependency strongly depends on the misalignment angle between the F1 and F2 layer magnetization $\theta$. 
The Gilbert damping magnitude decreases below the critical temperature for $\theta=0$ and $\theta=\pi/4$ while it increases for $\theta=\pi/2$ as shown in Fig. \ref{fig:fig_3} a). 
In this case, an additional damping torque appears below $T_c$ by the onset of equal-spin Cooper pairs.
The angle dependency of the Gilbert damping is shown in Fig. \ref{fig:fig_3} b). For $\theta\approx \pi/2$, the Gilbert damping is higher at zero temperature, $\delta G/G_{T>T_c}>0$ while it is smaller for $\theta < \theta^c$ where $\theta^c$ is the angle where $\delta G$ changes its sign (see Fig. \ref{fig:fig_3} b). The Gilbert damping is weakly affected by the S layer thickness and becomes constant for $d_S>\xi_0$ as shown in Fig. \ref{fig:fig_2} b). Nevertheless, the Gilbert damping is affected by the inverse proximity effect and the value of $\theta^c$ depends on the quality of the interfaces (see Fig. \ref{fig:fig_3} c). The generation process for equal-spin Cooper pairs is affected by the value of the superconducting gap at the F2/S interface which depends on the conductivity ratio $\sigma_S/\sigma_{F2}$. 
\begin{figure}
    \centering
    \includegraphics[width=8.6cm]{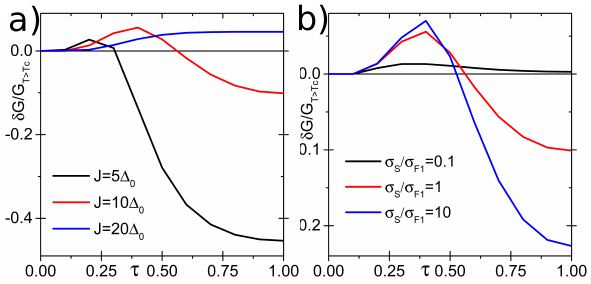}
    \caption{(Color online) a) Gilbert damping parameter $\delta G/G_{T>T_c}$ as a function of the interfaces transparencies $\tau $ for various exchange field magnitudes and $\theta=\pi/2$. b) Gilbert damping parameter $\delta G/G_{T>T_c}$ as a function of the interface transparencies $\tau $ for exchange field $J=10 \Delta_0$ and various S layer conductivities, and $\theta=\pi/2$. The other parameters are the same as in Fig. \ref{fig:fig_2}}
    \label{fig:fig_4}
\end{figure}

The Gilbert damping parameter $\delta G/G_{T>T_c}$ at $\theta=\pi/2$ depends non-monotonically on the interfaces transparencies and exchange field amplitude as shown in Fig. \ref{fig:fig_4} a). For small transparencies and exchange field $J\leq10\Delta_0$, the Gilbert damping amplitude is increased below the critical temperature $\delta G/G_{T>T_c}>0$ while it decreases $\delta G/G_{T>T_c}<0$ for transparencies close to 1 (see Fig. \ref{fig:fig_4} a). For higher exchange field amplitude, $J=20\Delta_0$, the injected spin current is increased for all transparencies. This non-monotonic behavior depends on the inverse proximity effect as shown in Fig. \ref{fig:fig_4} b). For an exchange field $J=10\Delta_0$, the Gilbert damping is reduced for transparency close to 1 when the inverse proximity effect is weak $\sigma_{S}/\sigma_{F1}=10$ while it increases for strong inverse proximity effect $\sigma_{S}/\sigma_{F1}=0.1$

\textit{Discussion}
In an S/F/S Josephson junction, the precession of the F layer magnetisation should produce equal-spin Cooper pairs originating the misalignment between the non-equilibrium magnetization and the F layer magnetization \cite{Houzet_PRL101_2008}. This effect strongly depends on the misalignment angle value and must be negligible when it tends to zero. In our calculation, this effect is negligible since $\theta_{F1} \rightarrow 0$.

The behavior of the Gilbert damping below $T_c$ can be understood as a competition between a decrease originating from standard Andreev reflection processes \cite{Bell_PRL2008,Morten_EPL2008} and spin-flip Andreev reflection processes \cite{Grein_PRB2010,Zutic_PRL2015}. In the latter, one electron is transmitted in the S layer as an equal-spin Cooper pair while a hole with the same spin is retro-reflected \cite{Grein_PRB2010,Zutic_PRL2015}. We expect this process to exist at interfaces with SOC \cite{Zutic_PRL2015} or spin-polarization \cite{Grein_PRB2010,Eschrig_NJPhys2015}.
For small angle,$\theta<\theta^c$, the standard Andreev reflection process dominates leading to a decrease of the injected spin current while for  $\theta>\theta^c$ the spin-flip Andreev reflection process is dominant.

Experimentally, we expect this effect to be observable in interfaces with inhomogeneous spin-polarisation as in Fe/Cr interfaces \cite{Robinson_PRB2014,Ishikawa_JPCS1965,Strom_Olsen_1979,Babic_JPCS1980}.

{\color{black} In our calculation, the precession of the F2 layer magnetization implies that spin may be injected back to the F1 layer reducing the Gilbert damping. We expect that in the presence of a static F2 layer magnetization, the Gilbert damping increase below Tc should be even higher.}

{\color{black} We can estimate the Gilbert damping from the calculated spin current density. In equation (\ref{Eq_spincurrent}), the spin current density is $I_s^0 \approx 10^{10} eV cm^-2$ considering a total sample thickness of $2.15 \xi_0$, a Fermi level density of states $N_0 \approx 10^{22} eV^{-1} cm^{-3}$ and a diffusive coefficient $D=\frac{\xi_0^2 \Delta_0}{\hbar}$ with $\Delta_0=1.4 meV$ the bulk Niobium superconducting gap at zero temperature. In Eq.(\ref{Gt}), the proportionality factor is $\beta=\frac{\gamma_0}{Ms d_{F1} f_{res}}$ \cite{Simensen_PRB2021} with $\gamma_0$ the gyromagnetic ratio and $Ms$ the magnetization saturation. Considering $\gamma_0\approx 10^7s^{-1}T^{-1}$ and $Ms \approx 10^{14} eV T^{-1} cm^{-3}$, we obtain $\beta \approx 1.6\times 10^{-11} eV^{-1} cm^2$. From the value of the current in the F1 layer plotted in the Fig. \ref{fig:fig_2}, we can estimate the Gilbert damping produced by the injected spin current in Eq.(\ref{Gt}): $G=\beta Is^Z$ and we find $G^{T=0} \approx 0.007$ at zero temperature and $G^{T>T_c} \approx 0.01$ above Tc. These values are of the same order of magnitude as the ones measured in Pt/Nb/Py/Nb/Pt \cite{Jeon_NatMat2018} and Nb/Cr/Fe/Cr/Nb \cite{Chan_arxiv2022} pentalayers.}

\textit{Conclusion}
We provide the first theoretical evidence that equal-spin Cooper pairs can enhance the injected spin current in superconducting nanostructures below the superconducting critical temperature. We anticipate that such proof will play a crucial role in the interpretation of forthcoming experiments and will influence further developments for applications in the field of superconducting spintronics.

\appendix

\section{Green function and Self-energy structure in the Keldysh formalism} \label{appA}
\subsection{Green function and self-energy general structure in Keldysh space}

Green's functions in quasiclassical theory of superconductivity exhibit internal and external degrees of freedom. The external degrees of freedom describe the motion in space (either ballistic along quasiclassical trajectories, or diffusive in the case of Usadel theory), while the internal degrees of freedom (2$\times $2 spin and 2$\times $2 particle-hole degrees of freedom) are discrete in nature and are represented by a matrix structure of the Green's function. In non-equilibrium the powerful Keldysh formalism adds a further 2$\times $2 matrix structure.
In the 8$\times$8 spin$\times$particle-hole$\times$Keldysh space, the Green's functions exhibit the following overall matrix structures \cite{Eschrig_adv2004,Eschrig_PRB2009}:
\begin{equation}
\begin{array}{ccccc}
\check{G}=\left(\begin{array}{cc}
\hat{G}^{R} & \hat{G}^{K}\\
0 & \hat{G}^{A}
\end{array}\right) & \hat{G}^{R,A}=\left(\begin{array}{cc}
g^{R,A} & f^{R,A}\\
\widetilde{f}^{R,A} & \widetilde{g}^{R,A}
\end{array}\right) \\
\hat{G}^{K}=\left(\begin{array}{cc}
g^{K} & f^{K}\\
-\widetilde{f}^{K} & -\widetilde{g}^{K}
\end{array}\right)\\
\end{array}\label{eq:Gsym}
\end{equation}
where $\check{...}$ corresponds to functions written
in the full $8\times8$ Keldysh$\otimes$spin$\otimes$particle-hole space ($\otimes$ is the tensorial product), and $\hat{...}$ corresponds to the 4$\times$4 spin$\times$particle-hole
space. The symbol $\widetilde{...}$ combines a complex
conjugation with the change $E\rightarrow-E$ in the energy argument, i.e., $\tilde{f}\left(E\right)=f^{*}\left(-E\right)$.
In the quasiclassical approach, both in the ballistic regime (Eilenberger formalism) and in the diffusive regime (Usadel formalism), the Green's function fulfills the normalization condition :
\begin{equation}
\check{G}.\check{G}=-\pi^{2}\check{1} .
\label{norm}
\end{equation}
In both Eilenberger and Usadel equations, there appear self-energies whose matrix structure in Keldysh space is given by
\begin{equation}
\begin{array}{ccccc}
\check{\Sigma}=\left(\begin{array}{cc}
\hat{\Sigma}^{R} & \hat{\Sigma}^{K}\\
0 & \hat{\Sigma}^{A}
\end{array}\right) &  \hat{\Sigma}^{R,A}=\left(\begin{array}{cc}
\Sigma^{R,A} & \Delta^{R,A}\\
\widetilde{\Delta}^{R,A} & \widetilde{\Sigma}^{R,A}
\end{array}\right) \\
\hat{\Sigma}^{K}=\left(\begin{array}{cc}
\Sigma^{K} & \Delta^{K}\\
-\widetilde{\Delta}^{K} & -\widetilde{\Sigma}^{K}
\end{array}\right)\\
\end{array}\label{eq:selfsym}
\end{equation}
where $\check{\Sigma}$ is the self energy written in the full $8\times8$ Keldysh$\otimes$spin$\otimes$particle times space, while $\hat{\Sigma}$ are self-energies written in the spin$\otimes$particle-hole space. $\Sigma$ and $\tilde{\Sigma}$ are normal self-energies in the 2$\times$2 spin space while $\Delta$ and $\tilde{\Delta}$ are anomalous self-energies in the spin space.
In the following, we refer to the $2\times2$ spin space through the unit matrix $\sigma_0$ and the three spin Pauli matrices $\left(\sigma^X,\sigma^Y,\sigma^Z\right)$, while we refer to the 2$\times$2 particle-hole space via the unit matrix $\tau_0$ and the three Pauli matrices $\left(\tau_1,\tau_2,\tau_3\right)$.

\subsection{Self energies in the Usadel equation \ref{eq_us}}

The self-energy $\check{\Sigma}$ including the single-particle terms describing the exchange splitting of spin bands in the ferromagnet writes in the full Keldysh space
\[
\check{\Sigma}=\check{\Sigma}^{imp}+\check{\Sigma}^{ex}+\check{\Delta}
\]
where $\check{\Sigma}^{imp}=\left(\begin{array}{cc}
\hat{\Sigma}^{imp,R} & \hat{\Sigma}^{imp,K}\\
0 & \hat{\Sigma}^{imp,A}
\end{array}\right)$ is the self-energy produced by the spin-flip on magnetic impurities
and spin-orbit scattering, $\check{\Sigma}^{ex}=\left(\begin{array}{cc}
\hat{\Sigma}^{ex,R} & \hat{\Sigma}^{ex,K}\\
0 & \hat{\Sigma}^{ex,A}
\end{array}\right)$ describes the spin-splitting of the energy bands produced by the exchange field in the F layer,
and $\check{\Delta}=\left(\begin{array}{cc}
\hat{\Delta}^{R} & \hat{\Delta}^{K}\\
0 & \hat{\Delta}^{A}
\end{array}\right)$ is the self energy describing singlet superconductivity. Note that $\hat{\Sigma}^{ex,K}=\hat{\Delta}^{K}=0$ while in general $\hat{\Sigma}^{imp,K}\neq0$.
In the following, we work only in the retarded subspace where $\hat{\Sigma}^{R}=\hat{\Sigma}^{imp,R}+\hat{\Sigma}^{ex,R}+\hat{\Delta}^{R}$ with
\[
\begin{array}{ccccc}
\hat{\Sigma}^{imp,R}=\left(\begin{array}{cc}
\Sigma^{imp} & \Delta^{imp}\\
\widetilde{\Delta}^{imp} & \widetilde{\Sigma}^{imp}
\end{array}\right) &  & \hat{\Sigma}^{ex,R}=\left(\begin{array}{cc}
\Sigma^{ex} & 0\\
0 & \widetilde{\Sigma}^{ex}
\end{array}\right) \\
\hat{\Delta}^{R}=\left(\begin{array}{cc}
0 & \Delta^{SC}\\
\widetilde{\Delta}^{SC} & 0
\end{array}\right)\end{array}
\]

\subsubsection{Spin-flip impurities self-energy}

\paragraph{Spin flip due to magnetic impurities}

The spin-flip scattering self-energy due to magnetic impurities writes
in the Keldysh space as
\[
\check{\Sigma}^{m}=\frac{1}{8\tau_{m}}\hat{\boldsymbol{\sigma}}.\check{1}.\check{G}.\hat{\boldsymbol{\sigma}}.\check{1}
\]
where $\hat{\boldsymbol{\sigma}}$ is the spin Pauli matrice vector in the
Nambu-spin space $\hat{\boldsymbol{\sigma}}=\left(\begin{array}{cc}
\boldsymbol{\sigma} & 0\\
0 & \boldsymbol{\sigma}*
\end{array}\right)$ .

\paragraph{Spin flip due to spin-orbit scattering}

The spin-orbit scattering self-energy writes in the Keldysh space
as
\[
\check{\Sigma}^{SO}=\frac{1}{8\tau_{so}}\hat{\boldsymbol{\sigma}}.\check{1}.\hat{\tau}_{3}.\check{G}.\hat{\tau}_{3}.\hat{\boldsymbol{\sigma}}.\check{1}.
\]

\subsubsection{Exchange field}

In the ferromagnet, we consider an exchange field oriented along the
$z$ axis which can be included by formally introducing a
self-energy term as
\[
\hat{\Sigma}^{ex,R}=\left(\begin{array}{cc}
\boldsymbol{J}\boldsymbol{\sigma} & 0\\
0 & \boldsymbol{J}\boldsymbol{\sigma}^{*}
\end{array}\right)
\]
where $\boldsymbol{J}$ is the exchange field and $\boldsymbol{\sigma}$
is the vector of Pauli matrices.

The symmetries between advanced and retarded functions are $\hat{\Sigma}^{ex,A}=\left(\widetilde{\varSigma}^{ex,R}\right)^{\dagger}=\left(\begin{array}{cc}
\boldsymbol{J}\boldsymbol{\sigma} & 0\\
0 & \boldsymbol{J}\boldsymbol{\sigma}^{*}
\end{array}\right)=\hat{\Sigma}^{ex}$.

\subsubsection{Superconducting self-energy}

In a spin-singlet superconductor, the order parameter is given by
\[
\hat{\Delta}^{R}=\left(\begin{array}{cc}
0 & \Delta^{SC}\\
\widetilde{\Delta}^{SC} & 0
\end{array}\right)
\]
where $\Delta^{SC}=i\sigma_{y}\Delta e^{i\phi}$ with $\phi$ the
superconducting phase.

The symmetries between advanced and retarded functions are $\Delta^{A}=-\left(\widetilde{\Delta}^{R}\right)^{\dagger}=\left(\begin{array}{cc}
0 & \Delta^{SC}\\
\widetilde{\Delta}^{SC} & 0
\end{array}\right)=\hat{\Delta}^{R}$. The singlet superconductivity order parameter is fixed by the self-consistency equation : 
\begin{equation}
\Delta\left(x\right)=\frac{\int_{-\infty}^{\infty}\frac{dE}{4i\pi}f_{s}^{K}\left(E,x\right)}{\int_{-\infty}^{\infty}\frac{dE}{2E}\text{tanh}\left(\frac{E}{2T}\right)+\text{ln}\left(\frac{T}{Tc}\right)}\label{eq:SC_self}
\end{equation}
where $f_s^K$ is the singlet part of the Keldysh anomalous Green function.

\section{Description of The FMR process with time dependent Usadel equations: From the laboratory frame to the rotating frame}\label{appB}  
\subsection{Time dependent Usadel equation}
The time-dependent Usadel equation in the Keldysh space is given by
\begin{equation}
i\left(\hat{\tau}_{3}\partial_{t_{1}}\check{G}+\partial_{t_{2}}\check{G}\hat{\tau}_{3}\right)+\frac{D}{\pi}\nabla_{\boldsymbol{R}}\left[\check{G}\circ\nabla_{\boldsymbol{R}}\left[\check{G}\right]\right]-\left[\check{\Sigma},\circ \check{G}\right]=0
\label{eq_us_t}
\end{equation}
where $\check{G}=\check{G}\left(t_1,t_2,\mathbf{R}\right)$ with $t_1$ and $t_2$ the time coordinates and $\mathbf{R}$ the space coordinate. The symbol $\circ$ denotes the time convolution product defined as :
\[
A\circ B (t_1,t_2)=\int_{-\infty}^{\infty}dt'A(t_1,t')B(t',t_2)
\]
Solution of Usadel's equation with a time dependence is complicated by the evaluation of these time convolution products. Note that in the above equation, the self-energy $\check{\Sigma}$ is time-dependent too. For our purpose it is sufficient to consider the case where $\check{\Sigma}\left(t_1,t_2\right)=\check{\Sigma}\left(t_1\right)\delta\left(t_1-t'\right)$.
\subsection{Exchange field time dependency in the laboratory frame}
Considering the ferromagnetic resonance process, we must take into account the time dependency of the F layer exchange field. Close to the resonance, we consider that the magnetization precesses around an effective field direction. Assuming an effective field directed along the $z$-axis, the time-dependency of the exchange field is given by
\begin{equation}
\mathbf{h}\left(t\right)=|h|\left(sin\left(\theta\right)sin\left(\Omega t\right),sin\left(\theta\right)cos\left(\Omega t\right),cos\left(\theta\right)\right)
    \label{h_dep_t}
\end{equation}
where $t$ is the time, $\theta$ is the tilting angle from the $z$ axis and $\Omega$ is the precession frequency. Note that for $t=0$, the magnetization is tilted from the $z$ axis towards the $y$ axis by the angle $\theta$.
In the Usadel's equation (\ref{eq_us_t}), a term due to the exchange field is present. One can separate the corresponding self-energy term $\check{\Sigma}\left(t\right)$ in a time-dependent and time-independent part $\check{\Sigma}\left(t\right)=\check{h}\left(t\right)+\check{\Sigma}_0$ where $\check{h}\left(t\right)$ is the self-energy associated with the exchange field. The self-energy $\check{h}$ exhibits the symmetry in the Keldysh space :
\begin{equation}
\begin{array}{ccc}
\check{h}=\left(\begin{array}{cc}
\hat{h} & 0\\
0 & \hat{h}
\end{array}\right) & , & \hat{h}=\left(\begin{array}{cc}
\mathbf{h\sigma} & 0\\
0 & \mathbf{h\sigma^*}
\end{array}\right) 
\end{array}
\label{h_self}
\end{equation}
where $\mathbf{\sigma}$ is the Pauli matrix vector in spin space. The Usadel's equation (\ref{eq_us_t}) then reads
\begin{equation}
\begin{array}{c}
i\left(\hat{\tau}_{3}\partial_{t_{1}}\check{G}+\partial_{t_{2}}\check{G}\hat{\tau}_{3}\right)+\frac{D}{\pi}\circ\nabla_{\boldsymbol{R}}\left[\check{G}\nabla_{\boldsymbol{R}}\left[\check{G}\right]\right]\\
-\left[\check{h}\left(t\right),\circ \check{G}\right]-\left[\check{\Sigma}_0, \check{G}\right]=0
\end{array}
\label{eq_us_t2}
\end{equation}
Note that the convolution product disappears from the last term of the equation because the self-energy $\check{\Sigma}_0$ is time-independent.
\subsection{From the laboratory frame to the rotating frame}
Assuming the time-dependency of the exchange field described in Eq.~(\ref{h_dep_t}), we define a unitary transformation which transforms the time-dependent Usadel equation in the laboratory frame to a time-independent Usadel equation in the rotating frame. This transformation is possible only if the exchange field exhibits the time dependency shown in the formula (\ref{h_dep_t}). For another time-dependency, one must solve the time dependent Usadel equation.
We can define this unitary transformation through the unitary operator $\check{U}$ which has the following structure in the Keldysh space,
\begin{equation}
\begin{array}{ccc}
\check{U}\left(t\right)=\left(\begin{array}{cc}
\hat{U}\left(t\right) & 0\\
0 & \hat{U^{\dagger}}\left(t\right)
\end{array}\right) & , & \hat{U}=\left(\begin{array}{cc}
U & 0\\
0 & U^{*}
\end{array}\right) 
\end{array}
\end{equation}
with the operator $U=e^{-i\sigma_Z \frac{\Omega t}{2}}$ where $\sigma_Z$ is the third Pauli matrix. In spin space, the transformation operator is given by $U=cos\left(\frac{\Omega t}{2}\right)-i \sigma_Z sin\left(\frac{\Omega t}{2}\right)$. Unitarity of the transformation imposes that $\check{U}\left(t_1\right)\check{U}^{\dagger}\left(t_2\right)=\delta \left(t_2-t_1\right)$ where $\delta$ is the Dirac distribution.
From this transformation, one can relate the Green's function in the rotating frame $\bar{\check{G}}$ to the Green's function in the laboratory frame $\check{G}$ via the transformation $\check{U}$ by
\begin{equation}
\bar{\check{G}}\left(t_1,t_2,\mathbf{R}\right)=\check{U}\left(t_1\right)\check{G}\left(t_1,t_2,\mathbf{R}\right)\check{U}^{\dagger}\left(t_2\right)
\end{equation}
Applying this transformation to the Usadel equation (\ref{eq_us_t2}) (multiplying on the left by $\check{U}\left(t_1\right)$ then on the right by $\check{U}^{\dagger}\left(t_2\right)$ and considering the unitary relation of $\check{U}$), we find
\begin{equation}
\begin{array}{c}
i\left(\left[\partial_{t_{2}}\overline{\check{G}}\hat{\tau}_{3}\check{1}+\hat{\tau}_{3}\check{1}\partial_{t_{1}}\overline{\check{G}}\right]\right)+\frac{D}{\pi}\nabla_{\boldsymbol{R}}\left[\bar{\check{G}}\circ\nabla_{\boldsymbol{R}}\left[\bar{\check{G}}\right]\right]\\
-\left[\frac{\Omega}{2}\sigma_{z}\check{1},\overline{\check{G}}\right]-\left[\check{h}_{eff},\bar{\check{G}}\right]-\left[\check{\Sigma}_0,\bar{\check{G}}\right]=0
\end{array}
\label{eq_us_rot_t}
\end{equation}
where $\check{h}_{eff}=\check{U}\check{h}\check{U}^{\dagger}$ is the exchange field in the rotating frame with the same structure as described in Eq.(\ref{h_self}) with a time-independent exchange field $\mathbf{h}_{eff}=|h|\left(0,sin\left(\theta\right),cos\left(\theta\right)\right)$. Working in the rotating frame imposes an additional exchange field along the $z$ direction the intensity of which is proportional to $\Omega/2$. This additional term is produced by the transformation of the time derivative term of Eq.(\ref{eq_us_t}).
This term reads
\[\check{U}\left(t_1\right)\left[i\left(\hat{\tau}_{3}\partial_{t_{1}}\check{G}+\partial_{t_{2}}\check{G}\hat{\tau}_{3}\right)\right]\check{U}^{\dagger}\left(t_2\right)
\]
and reduces to
\[i\left(\hat{\tau}_{3}\partial_{t_{1}}\bar{\check{G}}+\partial_{t_{2}}\bar{\check{G}}\hat{\tau}_{3}\right)-\left[\frac{\Omega}{2}\sigma_{z}\check{1},\bar{\check{G}} \right] .
\]
The source term in the Usadel equation (\ref{eq_us_rot_t}) (the commutator term) does not depend on time which implies that the Green's function only depends on the time difference $\delta t = t_1-t_2$, $\bar{\check{G}}\left(t_1,t_2,\mathbf{R}\right)=\bar{\check{G}}\left(\delta t,\mathbf{R}\right)$.
We then consider the Fourier transform:
\[
\check{G}\left(\delta t,\mathbf{R}\right)=\int dE G\left(E,\mathbf{R}\right)e^{iE\delta t}.
\]
Applying this Fourier transform to the Usadel equation (\ref{eq_us_rot_t}), we find
\begin{equation}
\frac{D}{\pi}\nabla_{\boldsymbol{R}}\left[\bar{\check{G}}\nabla_{\boldsymbol{R}}\left[\bar{\check{G}}\right]\right]+\left[E\hat{\tau}_{3}-\frac{\Omega}{2}\sigma_{z}\check{1}-\check{h}_{eff}-\check{\Sigma}_0,\bar{\check{G}}\right]=0 .
\label{eq_us_rot_E}
\end{equation}
This time-independent Usadel equation (\ref{eq_us_rot_E}) describes the superconducting physics in the rotating frame. In order to return back to the description in the laboratory frame, one has to apply the inverse transformation onto the Green's function.

\section{Usadel's equations in the gamma parametrization}\label{appC}  
Here we present the Usadel equation for the matrices $\gamma$ and $\tilde{\gamma}$. We start with the Usadel equation
\begin{equation}
D\nabla_{\boldsymbol{R}}\left[\check{g}\nabla_{\boldsymbol{R}}\left[\check{g}\right]\right]+i\left[E\hat{\tau}_{3}-\check{\Sigma},\check{g}\right]=0
\label{eq_us_ti}
\end{equation}
which is the same than the equation (\ref{eq_us_rot_E}), where for simplicity we write $\bar{\check{G}}=\check{G}$, $\check{G}=-i\pi\check{g}$ and $\check{\Sigma}=\frac{\Omega}{2}\sigma_{z}\check{1}+\check{h}_{eff}+\check{\Sigma}_0$. The Usadel equation (\ref{eq_us_ti}) is divided in two distinct terms : the spatial derivative term $\nabla_{\boldsymbol{R}}\left[\check{g}\nabla_{\boldsymbol{R}}\left[\check{g}\right]\right]$ and the non-derivative term $\left[\check{\Sigma},\check{g}\right]$. It is convenient parameterize the Green's function such that it already fulfills the normalisation condition (\ref{norm}). Here, we use the Riccatti matrix parametrization \cite{Eschrig_PRB2009} where the Green's functions are given in  4$\times$4 spin$\times$particle-hole space by
\begin{equation}
 \hat{G}^{K}=-2i\pi.\hat{N}^{R}.\left(\begin{array}{cc}
\left(x-\gamma^{R}.\tilde{x}.\widetilde{\gamma}^{A}\right) & -\left(\gamma^{R}.\widetilde{x}-x.\gamma^{A}\right)\\
-\left(\widetilde{\gamma}^{R}.x-\widetilde{x}.\widetilde{\gamma}^{A}\right) & \left(\tilde{x}-\widetilde{\gamma}^{R}.x.\gamma^{A}\right)
\end{array}\right).\hat{N}^{A}   
\label{gk_gam}
\end{equation}
and 
\begin{equation}
   \hat{G}^{R,A}=\mp i\pi.\hat{N}^{R,A}.\left(\begin{array}{cc}
1+\gamma^{R,A}\widetilde{\gamma}^{R,A} & 2\gamma^{R,A}\\
-2\widetilde{\gamma}^{R,A} & -\left(1+\widetilde{\gamma}^{R,A}\gamma^{R,A}\right)
\end{array}\right) .
\label{Gr_gam}
\end{equation}

\subsection{Usadel's equation for retarded Green's functions}
In this section, we focus on the Usadel equation for retarded component which reads
\begin{equation}
D\nabla\left[\hat{g}^R\nabla\left[\hat{g}^R\right]\right]+i\left[E\hat{\tau}_{3}-\hat{\Sigma}^R,\hat{g}^R\right]=0 .
\label{eq_us_tir}
\end{equation}
It leads to the equation
\begin{equation}
\begin{array}{c}
D\left[\nabla^{2}\gamma^{R}+2\nabla\gamma^{R}\widetilde{N}^{R}\widetilde{\gamma}^{R}\nabla\gamma^{R}\right]\\
+i\left(2E\gamma^{R}-\Sigma^{R}\gamma^{R}+\gamma^{R}\widetilde{\Sigma}^{R}+\Delta^{R}-\gamma^{R}\widetilde{\Delta}^{R}\gamma^{R}\right)=0
\end{array}
\label{eq_gamr}
\end{equation}
which corresponds to a differential equations for $\gamma^R$.  
Solving the equations for $\gamma^R$ and $\tilde{\gamma}^R$ can be achieved by numerical methods as for example relaxation methods. From the solution, one uses Eq. (\ref{Gr_gam}) to build the retarded Green's function. Using the symmetries between retarded and advanced Green's functions described in Eq.(\ref{Sym_gamma}), we can derive the expression for $\gamma^A$ and $\tilde{\gamma}^A$ and construct the advanced Green's function.
\subsection{Usadel's equation for Keldysh Green's function: quantum kinetic equation}
To calculate the properties of diffusive superconducting nanostructures, we consider the equation for non-equilibrium distribution functions i.e. the quantum kinetic equation. In this section, we derive the kinetic equation for the distribution function $x$ and $\tilde{x}$ from Eqs.(\ref{gk_gam}).

\subsubsection{Distribution function in the Riccatti matrices parametrization}
The choice for the the distribution functions $x$ and $\tilde{x}$ is not unique \cite{Eschrig_PRB2009}. The choice of the functions $x$ and $\tilde{x}$ leads to a simplification of the kinetic equations. These distribution functions are related to the distribution function $h$ and $\tilde{h}$ introduced by Larkin and Ovchinikov as \cite{Eschrig_PRB2009}  by
\[
h=\sum_{n=0}^{n=\infty}\left[\left(\gamma^R\tilde{\gamma}^R\right)^n\circ\left(x-\gamma^R\circ \tilde{x}\circ\tilde{\gamma}^A\right)\circ\left(\gamma^A\tilde{\gamma}^A\right)^n\right]
\]
where $\tilde{h}$ can be deduced by applying the $\tilde{...}$ transformation to the function $h$. The distribution function $h$ can be related to the distribution functions for electrons and holes $f$ and $\tilde{f}$ by the relation 
\[
\begin{array}{c}
f=\frac{1}{2}\left(1-h\right)\\
\tilde{f}=\frac{1}{2}\left(1+\tilde{h}\right)
\end{array}
\]
From the resolution of the kinetic equation, we can calculate the space dependency of the distribution function in the system.

\subsubsection{The kinetic equations for the distribution functions $x$  and $\tilde{x}$}
We derive the kinetic equations from the Keldysh part of the Usadel equation (\ref{eq_us_rot_E}).
\paragraph{Full kinetic equation}
We find the kinetic equation in the form \cite{Eschrig_adv2004,Eschrig_PRB2009}:
\begin{equation}
\begin{array}{c}
D\left\{ \nabla^{2}x^{K}+2\nabla\gamma^{R}\widetilde{N}^{R}\widetilde{\gamma}^{R}\nabla x^{K}+2\nabla x^{K}N^{A}\gamma^{A}\nabla\widetilde{\gamma}^{A}\right.\\
\left.-2\nabla\gamma^{R}\widetilde{N}^{R}\left[\widetilde{x}^{K}-\widetilde{\gamma}^{R}x^{K}\gamma^{A}\right]\widetilde{N}^{A}\nabla\widetilde{\gamma}^{A}\right\} \\
+i\left(-\Sigma^{R}-\gamma^{R}\widetilde{\Delta}^{R}\right)x^{K}
+ix^{K}\left(\Sigma^{A}-\Delta^{A}\widetilde{\gamma}^{A}\right)\\
+i\left(\Sigma^{K}-\Delta^{K}\widetilde{\gamma}^{A}-\gamma^{R}\widetilde{\Delta}^{K}+\gamma^{R}\widetilde{\Sigma}^{K}\widetilde{\gamma}^{A}\right)=0
\end{array}
\label{eq:kinetic_xk}
\end{equation}
The kinetic equation for $\tilde{x}^K$ can be obtained by applying the projection $\hat{P}_{-}^{R}(...)\hat{P}_{+}^{A}$ of the Usadel equation or applying the $\tilde{...}$ transformation to Eq. (\ref{eq:kinetic_xk}).

\section{Symmetries in Keldysh-space}\label{appD}
The following symmetries connect retarded components with advanced components and express symmetries of Keldysh components: \cite{Eschrig_PRB2009} :
\begin{equation}
\begin{array}{ccccc}
\gamma^{A}=\left(\widetilde{\gamma}^{R}\right)^{\dagger} &  & \Delta^{A}=-\left(\widetilde{\Delta}^{R}\right)^{\dagger} &  & \varSigma^{A}=\left(\widetilde{\varSigma}^{R}\right)^{\dagger}\\
\\
x=\left(x\right)^{\dagger} &  & \Delta^{K}=\left(\widetilde{\Delta}^{K}\right)^{\dagger} &  & \varSigma^{K}=-\left(\varSigma^{K}\right)^{\dagger}
\end{array}
\label{Sym_gamma}
\end{equation}

\begin{acknowledgments}
The  authors acknowledge M. Blamire, A. K. Chan, L. Cohen, K.-R. Jeon, H. Kurebayashi, and J. Robinson for fruitful discussions. This work has been supported by the EPSRC Programme Grant EP/N017242/1.
\end{acknowledgments}

\bibliography{Supra}

\begin{thebibliography}{70}%
\makeatletter
\providecommand \@ifxundefined [1]{%
 \@ifx{#1\undefined}
}%
\providecommand \@ifnum [1]{%
 \ifnum #1\expandafter \@firstoftwo
 \else \expandafter \@secondoftwo
 \fi
}%
\providecommand \@ifx [1]{%
 \ifx #1\expandafter \@firstoftwo
 \else \expandafter \@secondoftwo
 \fi
}%
\providecommand \natexlab [1]{#1}%
\providecommand \enquote  [1]{``#1''}%
\providecommand \bibnamefont  [1]{#1}%
\providecommand \bibfnamefont [1]{#1}%
\providecommand \citenamefont [1]{#1}%
\providecommand \href@noop [0]{\@secondoftwo}%
\providecommand \href [0]{\begingroup \@sanitize@url \@href}%
\providecommand \@href[1]{\@@startlink{#1}\@@href}%
\providecommand \@@href[1]{\endgroup#1\@@endlink}%
\providecommand \@sanitize@url [0]{\catcode `\\12\catcode `\$12\catcode
  `\&12\catcode `\#12\catcode `\^12\catcode `\_12\catcode `\%12\relax}%
\providecommand \@@startlink[1]{}%
\providecommand \@@endlink[0]{}%
\providecommand \url  [0]{\begingroup\@sanitize@url \@url }%
\providecommand \@url [1]{\endgroup\@href {#1}{\urlprefix }}%
\providecommand \urlprefix  [0]{URL }%
\providecommand \Eprint [0]{\href }%
\providecommand \doibase [0]{http://dx.doi.org/}%
\providecommand \selectlanguage [0]{\@gobble}%
\providecommand \bibinfo  [0]{\@secondoftwo}%
\providecommand \bibfield  [0]{\@secondoftwo}%
\providecommand \translation [1]{[#1]}%
\providecommand \BibitemOpen [0]{}%
\providecommand \bibitemStop [0]{}%
\providecommand \bibitemNoStop [0]{.\EOS\space}%
\providecommand \EOS [0]{\spacefactor3000\relax}%
\providecommand \BibitemShut  [1]{\csname bibitem#1\endcsname}%
\let\auto@bib@innerbib\@empty
\bibitem [{\citenamefont {Eschrig}(2011)}]{Eschrig_PhysTod2011}%
  \BibitemOpen
  \bibfield  {author} {\bibinfo {author} {\bibfnamefont {M.}~\bibnamefont
  {Eschrig}},\ }\href {https://doi.org/10.1063/1.3541944} {\bibfield  {journal}
  {\bibinfo  {journal} {Phys. Today}\ }\textbf {\bibinfo {volume} {64}},\
  \bibinfo {pages} {43} (\bibinfo {year} {2011})}\BibitemShut {NoStop}%
\bibitem [{\citenamefont {Eschrig}(2015)}]{Eschrig_RepProgPhys2015}%
  \BibitemOpen
  \bibfield  {author} {\bibinfo {author} {\bibfnamefont {M.}~\bibnamefont
  {Eschrig}},\ }\href {http://stacks.iop.org/0034-4885/78/i=10/a=104501}
  {\bibfield  {journal} {\bibinfo  {journal} {Rep. Prog. Phys.}\ }\textbf
  {\bibinfo {volume} {78}},\ \bibinfo {pages} {104501} (\bibinfo {year}
  {2015})}\BibitemShut {NoStop}%
\bibitem [{\citenamefont {Linder}\ and\ \citenamefont
  {Robinson}(2015)}]{Robinson_Linder_2015}%
  \BibitemOpen
  \bibfield  {author} {\bibinfo {author} {\bibfnamefont {J.}~\bibnamefont
  {Linder}}\ and\ \bibinfo {author} {\bibfnamefont {J.~W.~A.}\ \bibnamefont
  {Robinson}},\ }\href {http://dx.doi.org/10.1038/nphys3242} {\bibfield
  {journal} {\bibinfo  {journal} {Nat. Phys.}\ }\textbf {\bibinfo {volume}
  {11}},\ \bibinfo {pages} {307} (\bibinfo {year} {2015})}\BibitemShut
  {NoStop}%
\bibitem [{\citenamefont {Izyumov}\ \emph {et~al.}(2002)\citenamefont
  {Izyumov}, \citenamefont {Proshin},\ and\ \citenamefont
  {Khusainov}}]{Izyumov2002}%
  \BibitemOpen
  \bibfield  {author} {\bibinfo {author} {\bibfnamefont {Y.~A.}\ \bibnamefont
  {Izyumov}}, \bibinfo {author} {\bibfnamefont {Y.~N.}\ \bibnamefont
  {Proshin}}, \ and\ \bibinfo {author} {\bibfnamefont {M.~G.}\ \bibnamefont
  {Khusainov}},\ }\href {http://stacks.iop.org/1063-7869/45/i=2/a=R01}
  {\bibfield  {journal} {\bibinfo  {journal} {Physics-Uspekhi}\ }\textbf
  {\bibinfo {volume} {45}},\ \bibinfo {pages} {109} (\bibinfo {year}
  {2002})}\BibitemShut {NoStop}%
\bibitem [{\citenamefont {Eschrig}\ \emph {et~al.}(2004)\citenamefont
  {Eschrig}, \citenamefont {Kopu}, \citenamefont {Konstandin}, \citenamefont
  {Cuevas}, \citenamefont {Fogelstr{\"o}m},\ and\ \citenamefont
  {Sch{\"o}n}}]{Eschrig_adv2004}%
  \BibitemOpen
  \bibfield  {author} {\bibinfo {author} {\bibfnamefont {M.}~\bibnamefont
  {Eschrig}}, \bibinfo {author} {\bibfnamefont {J.}~\bibnamefont {Kopu}},
  \bibinfo {author} {\bibfnamefont {A.}~\bibnamefont {Konstandin}}, \bibinfo
  {author} {\bibfnamefont {J.~C.}\ \bibnamefont {Cuevas}}, \bibinfo {author}
  {\bibfnamefont {M.}~\bibnamefont {Fogelstr{\"o}m}}, \ and\ \bibinfo {author}
  {\bibfnamefont {G.}~\bibnamefont {Sch{\"o}n}},\ }in\ \href@noop {} {\emph
  {\bibinfo {booktitle} {Advances in Solid State Physics}}}\ (\bibinfo
  {publisher} {Springer},\ \bibinfo {year} {2004})\ pp.\ \bibinfo {pages}
  {533--545}\BibitemShut {NoStop}%
\bibitem [{\citenamefont {Golubov}\ \emph {et~al.}(2004)\citenamefont
  {Golubov}, \citenamefont {Kupriyanov},\ and\ \citenamefont
  {Il'ichev}}]{Golubov_RevMod2004}%
  \BibitemOpen
  \bibfield  {author} {\bibinfo {author} {\bibfnamefont {A.~A.}\ \bibnamefont
  {Golubov}}, \bibinfo {author} {\bibfnamefont {M.~Y.}\ \bibnamefont
  {Kupriyanov}}, \ and\ \bibinfo {author} {\bibfnamefont {E.}~\bibnamefont
  {Il'ichev}},\ }\href {\doibase 10.1103/RevModPhys.76.411} {\bibfield
  {journal} {\bibinfo  {journal} {Rev. Mod. Phys.}\ }\textbf {\bibinfo {volume}
  {76}},\ \bibinfo {pages} {411} (\bibinfo {year} {2004})}\BibitemShut
  {NoStop}%
\bibitem [{\citenamefont {Bergeret}\ \emph {et~al.}(2005)\citenamefont
  {Bergeret}, \citenamefont {Volkov},\ and\ \citenamefont
  {Efetov}}]{Bergeret_RevMod2005}%
  \BibitemOpen
  \bibfield  {author} {\bibinfo {author} {\bibfnamefont {F.~S.}\ \bibnamefont
  {Bergeret}}, \bibinfo {author} {\bibfnamefont {A.~F.}\ \bibnamefont
  {Volkov}}, \ and\ \bibinfo {author} {\bibfnamefont {K.~B.}\ \bibnamefont
  {Efetov}},\ }\href {\doibase 10.1103/RevModPhys.77.1321} {\bibfield
  {journal} {\bibinfo  {journal} {Rev. Mod. Phys.}\ }\textbf {\bibinfo {volume}
  {77}},\ \bibinfo {pages} {1321} (\bibinfo {year} {2005})}\BibitemShut
  {NoStop}%
\bibitem [{\citenamefont {Buzdin}(2005)}]{Buzdin_RevMod2005}%
  \BibitemOpen
  \bibfield  {author} {\bibinfo {author} {\bibfnamefont {A.~I.}\ \bibnamefont
  {Buzdin}},\ }\href {\doibase 10.1103/RevModPhys.77.935} {\bibfield  {journal}
  {\bibinfo  {journal} {Rev. Mod. Phys.}\ }\textbf {\bibinfo {volume} {77}},\
  \bibinfo {pages} {935} (\bibinfo {year} {2005})}\BibitemShut {NoStop}%
\bibitem [{\citenamefont {Lyuksyutov}\ and\ \citenamefont
  {Pokrovsky}(2005)}]{Lyuksyutov2005}%
  \BibitemOpen
  \bibfield  {author} {\bibinfo {author} {\bibfnamefont {I.~F.}\ \bibnamefont
  {Lyuksyutov}}\ and\ \bibinfo {author} {\bibfnamefont {V.~L.}\ \bibnamefont
  {Pokrovsky}},\ }\href {\doibase 10.1080/00018730500057536} {\bibfield
  {journal} {\bibinfo  {journal} {Adv. Phys.}\ }\textbf {\bibinfo {volume}
  {54}},\ \bibinfo {pages} {67} (\bibinfo {year} {2005})}\BibitemShut {NoStop}%
\bibitem [{\citenamefont {Blamire}\ and\ \citenamefont
  {Robinson}(2014)}]{BlamireRobinson_JPhys2014}%
  \BibitemOpen
  \bibfield  {author} {\bibinfo {author} {\bibfnamefont {M.~G.}\ \bibnamefont
  {Blamire}}\ and\ \bibinfo {author} {\bibfnamefont {J.~W.~A.}\ \bibnamefont
  {Robinson}},\ }\href {http://stacks.iop.org/0953-8984/26/i=45/a=453201}
  {\bibfield  {journal} {\bibinfo  {journal} {J. Phys. Condens. Matter}\
  }\textbf {\bibinfo {volume} {26}},\ \bibinfo {pages} {453201} (\bibinfo
  {year} {2014})}\BibitemShut {NoStop}%
\bibitem [{\citenamefont {Birge}(2018)}]{Birge_PhilTransA2018}%
  \BibitemOpen
  \bibfield  {author} {\bibinfo {author} {\bibfnamefont {N.~O.}\ \bibnamefont
  {Birge}},\ }\href
  {http://rsta.royalsocietypublishing.org/content/376/2125/20150150} {\bibfield
   {journal} {\bibinfo  {journal} {Philos. Trans. Royal Soc. A}\ }\textbf
  {\bibinfo {volume} {376}} (\bibinfo {year} {2018})}\BibitemShut {NoStop}%
\bibitem [{\citenamefont {Tokuyasu}\ \emph {et~al.}(1988)\citenamefont
  {Tokuyasu}, \citenamefont {Sauls},\ and\ \citenamefont
  {Rainer}}]{Sauls_PRB1988}%
  \BibitemOpen
  \bibfield  {author} {\bibinfo {author} {\bibfnamefont {T.}~\bibnamefont
  {Tokuyasu}}, \bibinfo {author} {\bibfnamefont {J.~A.}\ \bibnamefont {Sauls}},
  \ and\ \bibinfo {author} {\bibfnamefont {D.}~\bibnamefont {Rainer}},\ }\href
  {\doibase 10.1103/PhysRevB.38.8823} {\bibfield  {journal} {\bibinfo
  {journal} {Phys. Rev. B}\ }\textbf {\bibinfo {volume} {38}},\ \bibinfo
  {pages} {8823} (\bibinfo {year} {1988})}\BibitemShut {NoStop}%
\bibitem [{\citenamefont {Fogelstr\"om}(2000)}]{Fogelstrom00}%
  \BibitemOpen
  \bibfield  {author} {\bibinfo {author} {\bibfnamefont {M.}~\bibnamefont
  {Fogelstr\"om}},\ }\href {\doibase 10.1103/PhysRevB.62.11812} {\bibfield
  {journal} {\bibinfo  {journal} {Phys. Rev. B}\ }\textbf {\bibinfo {volume}
  {62}},\ \bibinfo {pages} {11812} (\bibinfo {year} {2000})}\BibitemShut
  {NoStop}%
\bibitem [{\citenamefont {Barash}\ and\ \citenamefont
  {Bobkova}(2002)}]{BobkovaBobkov02}%
  \BibitemOpen
  \bibfield  {author} {\bibinfo {author} {\bibfnamefont {Y.~S.}\ \bibnamefont
  {Barash}}\ and\ \bibinfo {author} {\bibfnamefont {I.~V.}\ \bibnamefont
  {Bobkova}},\ }\href {\doibase 10.1103/PhysRevB.65.144502} {\bibfield
  {journal} {\bibinfo  {journal} {Phys. Rev. B}\ }\textbf {\bibinfo {volume}
  {65}},\ \bibinfo {pages} {144502} (\bibinfo {year} {2002})}\BibitemShut
  {NoStop}%
\bibitem [{\citenamefont {Eschrig}\ \emph {et~al.}(2003)\citenamefont
  {Eschrig}, \citenamefont {Kopu}, \citenamefont {Cuevas},\ and\ \citenamefont
  {Sch\"on}}]{Eschrig_PRL2003}%
  \BibitemOpen
  \bibfield  {author} {\bibinfo {author} {\bibfnamefont {M.}~\bibnamefont
  {Eschrig}}, \bibinfo {author} {\bibfnamefont {J.}~\bibnamefont {Kopu}},
  \bibinfo {author} {\bibfnamefont {J.~C.}\ \bibnamefont {Cuevas}}, \ and\
  \bibinfo {author} {\bibfnamefont {G.}~\bibnamefont {Sch\"on}},\ }\href
  {\doibase 10.1103/PhysRevLett.90.137003} {\bibfield  {journal} {\bibinfo
  {journal} {Phys. Rev. Lett.}\ }\textbf {\bibinfo {volume} {90}},\ \bibinfo
  {pages} {137003} (\bibinfo {year} {2003})}\BibitemShut {NoStop}%
\bibitem [{\citenamefont {Bergeret}\ \emph {et~al.}(2001)\citenamefont
  {Bergeret}, \citenamefont {Volkov},\ and\ \citenamefont
  {Efetov}}]{Bergeret_PRL2001}%
  \BibitemOpen
  \bibfield  {author} {\bibinfo {author} {\bibfnamefont {F.~S.}\ \bibnamefont
  {Bergeret}}, \bibinfo {author} {\bibfnamefont {A.~F.}\ \bibnamefont
  {Volkov}}, \ and\ \bibinfo {author} {\bibfnamefont {K.~B.}\ \bibnamefont
  {Efetov}},\ }\href {\doibase 10.1103/PhysRevLett.86.4096} {\bibfield
  {journal} {\bibinfo  {journal} {Phys. Rev. Lett.}\ }\textbf {\bibinfo
  {volume} {86}},\ \bibinfo {pages} {4096} (\bibinfo {year}
  {2001})}\BibitemShut {NoStop}%
\bibitem [{\citenamefont {L\"ofwander}\ \emph {et~al.}(2005)\citenamefont
  {L\"ofwander}, \citenamefont {Champel}, \citenamefont {Durst},\ and\
  \citenamefont {Eschrig}}]{Champel_PRL2005}%
  \BibitemOpen
  \bibfield  {author} {\bibinfo {author} {\bibfnamefont {T.}~\bibnamefont
  {L\"ofwander}}, \bibinfo {author} {\bibfnamefont {T.}~\bibnamefont
  {Champel}}, \bibinfo {author} {\bibfnamefont {J.}~\bibnamefont {Durst}}, \
  and\ \bibinfo {author} {\bibfnamefont {M.}~\bibnamefont {Eschrig}},\ }\href
  {\doibase 10.1103/PhysRevLett.95.187003} {\bibfield  {journal} {\bibinfo
  {journal} {Phys. Rev. Lett.}\ }\textbf {\bibinfo {volume} {95}},\ \bibinfo
  {pages} {187003} (\bibinfo {year} {2005})}\BibitemShut {NoStop}%
\bibitem [{\citenamefont {Houzet}\ and\ \citenamefont
  {Buzdin}(2007)}]{houzet_PRB2007}%
  \BibitemOpen
  \bibfield  {author} {\bibinfo {author} {\bibfnamefont {M.}~\bibnamefont
  {Houzet}}\ and\ \bibinfo {author} {\bibfnamefont {A.~I.}\ \bibnamefont
  {Buzdin}},\ }\href {\doibase 10.1103/PhysRevB.76.060504} {\bibfield
  {journal} {\bibinfo  {journal} {Phys. Rev. B}\ }\textbf {\bibinfo {volume}
  {76}},\ \bibinfo {pages} {060504} (\bibinfo {year} {2007})}\BibitemShut
  {NoStop}%
\bibitem [{\citenamefont {Halterman}\ \emph {et~al.}(2008)\citenamefont
  {Halterman}, \citenamefont {Valls},\ and\ \citenamefont
  {Barsic}}]{Halterman_PRB2008}%
  \BibitemOpen
  \bibfield  {author} {\bibinfo {author} {\bibfnamefont {K.}~\bibnamefont
  {Halterman}}, \bibinfo {author} {\bibfnamefont {O.~T.}\ \bibnamefont
  {Valls}}, \ and\ \bibinfo {author} {\bibfnamefont {P.~H.}\ \bibnamefont
  {Barsic}},\ }\href {\doibase 10.1103/PhysRevB.77.174511} {\bibfield
  {journal} {\bibinfo  {journal} {Phys. Rev. B}\ }\textbf {\bibinfo {volume}
  {77}},\ \bibinfo {pages} {174511} (\bibinfo {year} {2008})}\BibitemShut
  {NoStop}%
\bibitem [{\citenamefont {Champel}\ and\ \citenamefont
  {Eschrig}(2005{\natexlab{a}})}]{Champel_PRB2005}%
  \BibitemOpen
  \bibfield  {author} {\bibinfo {author} {\bibfnamefont {T.}~\bibnamefont
  {Champel}}\ and\ \bibinfo {author} {\bibfnamefont {M.}~\bibnamefont
  {Eschrig}},\ }\href {\doibase 10.1103/PhysRevB.71.220506} {\bibfield
  {journal} {\bibinfo  {journal} {Phys. Rev. B}\ }\textbf {\bibinfo {volume}
  {71}},\ \bibinfo {pages} {220506} (\bibinfo {year}
  {2005}{\natexlab{a}})}\BibitemShut {NoStop}%
\bibitem [{\citenamefont {Champel}\ and\ \citenamefont
  {Eschrig}(2005{\natexlab{b}})}]{Champel_PRB2005b}%
  \BibitemOpen
  \bibfield  {author} {\bibinfo {author} {\bibfnamefont {T.}~\bibnamefont
  {Champel}}\ and\ \bibinfo {author} {\bibfnamefont {M.}~\bibnamefont
  {Eschrig}},\ }\href {\doibase 10.1103/PhysRevB.72.054523} {\bibfield
  {journal} {\bibinfo  {journal} {Phys. Rev. B}\ }\textbf {\bibinfo {volume}
  {72}},\ \bibinfo {pages} {054523} (\bibinfo {year}
  {2005}{\natexlab{b}})}\BibitemShut {NoStop}%
\bibitem [{\citenamefont {Fominov}\ \emph {et~al.}(2007)\citenamefont
  {Fominov}, \citenamefont {Volkov},\ and\ \citenamefont
  {Efetov}}]{Fominov_PRB2007}%
  \BibitemOpen
  \bibfield  {author} {\bibinfo {author} {\bibfnamefont {Y.~V.}\ \bibnamefont
  {Fominov}}, \bibinfo {author} {\bibfnamefont {A.~F.}\ \bibnamefont {Volkov}},
  \ and\ \bibinfo {author} {\bibfnamefont {K.~B.}\ \bibnamefont {Efetov}},\
  }\href {\doibase 10.1103/PhysRevB.75.104509} {\bibfield  {journal} {\bibinfo
  {journal} {Phys. Rev. B}\ }\textbf {\bibinfo {volume} {75}},\ \bibinfo
  {pages} {104509} (\bibinfo {year} {2007})}\BibitemShut {NoStop}%
\bibitem [{\citenamefont {Crouzy}\ \emph {et~al.}(2007)\citenamefont {Crouzy},
  \citenamefont {Tollis},\ and\ \citenamefont {Ivanov}}]{Crouzy_PRB2007}%
  \BibitemOpen
  \bibfield  {author} {\bibinfo {author} {\bibfnamefont {B.}~\bibnamefont
  {Crouzy}}, \bibinfo {author} {\bibfnamefont {S.}~\bibnamefont {Tollis}}, \
  and\ \bibinfo {author} {\bibfnamefont {D.~A.}\ \bibnamefont {Ivanov}},\
  }\href {\doibase 10.1103/PhysRevB.76.134502} {\bibfield  {journal} {\bibinfo
  {journal} {Phys. Rev. B}\ }\textbf {\bibinfo {volume} {76}},\ \bibinfo
  {pages} {134502} (\bibinfo {year} {2007})}\BibitemShut {NoStop}%
\bibitem [{\citenamefont {Buzdin}\ \emph {et~al.}(2011)\citenamefont {Buzdin},
  \citenamefont {Mel'nikov},\ and\ \citenamefont {Pugach}}]{Pugach_PRB2011}%
  \BibitemOpen
  \bibfield  {author} {\bibinfo {author} {\bibfnamefont {A.~I.}\ \bibnamefont
  {Buzdin}}, \bibinfo {author} {\bibfnamefont {A.~S.}\ \bibnamefont
  {Mel'nikov}}, \ and\ \bibinfo {author} {\bibfnamefont {N.~G.}\ \bibnamefont
  {Pugach}},\ }\href {\doibase 10.1103/PhysRevB.83.144515} {\bibfield
  {journal} {\bibinfo  {journal} {Phys. Rev. B}\ }\textbf {\bibinfo {volume}
  {83}},\ \bibinfo {pages} {144515} (\bibinfo {year} {2011})}\BibitemShut
  {NoStop}%
\bibitem [{\citenamefont {Kupferschmidt}\ and\ \citenamefont
  {Brouwer}(2009)}]{Kupferschmidt_PRB2009}%
  \BibitemOpen
  \bibfield  {author} {\bibinfo {author} {\bibfnamefont {J.~N.}\ \bibnamefont
  {Kupferschmidt}}\ and\ \bibinfo {author} {\bibfnamefont {P.~W.}\ \bibnamefont
  {Brouwer}},\ }\href {\doibase 10.1103/PhysRevB.80.214537} {\bibfield
  {journal} {\bibinfo  {journal} {Phys. Rev. B}\ }\textbf {\bibinfo {volume}
  {80}},\ \bibinfo {pages} {214537} (\bibinfo {year} {2009})}\BibitemShut
  {NoStop}%
\bibitem [{\citenamefont {Eschrig}\ and\ \citenamefont
  {L{\"o}fwander}(2008)}]{eschrig_NatPhys2008}%
  \BibitemOpen
  \bibfield  {author} {\bibinfo {author} {\bibfnamefont {M.}~\bibnamefont
  {Eschrig}}\ and\ \bibinfo {author} {\bibfnamefont {T.}~\bibnamefont
  {L{\"o}fwander}},\ }\href {\doibase 10.1038/nphys831} {\bibfield  {journal}
  {\bibinfo  {journal} {Nat. Phys.}\ }\textbf {\bibinfo {volume} {4}},\
  \bibinfo {pages} {138} (\bibinfo {year} {2008})}\BibitemShut {NoStop}%
\bibitem [{\citenamefont {Annunziata}\ \emph {et~al.}(2012)\citenamefont
  {Annunziata}, \citenamefont {Manske},\ and\ \citenamefont
  {Linder}}]{Annunziata_PRB2012}%
  \BibitemOpen
  \bibfield  {author} {\bibinfo {author} {\bibfnamefont {G.}~\bibnamefont
  {Annunziata}}, \bibinfo {author} {\bibfnamefont {D.}~\bibnamefont {Manske}},
  \ and\ \bibinfo {author} {\bibfnamefont {J.}~\bibnamefont {Linder}},\ }\href
  {\doibase 10.1103/PhysRevB.86.174514} {\bibfield  {journal} {\bibinfo
  {journal} {Phys. Rev. B}\ }\textbf {\bibinfo {volume} {86}},\ \bibinfo
  {pages} {174514} (\bibinfo {year} {2012})}\BibitemShut {NoStop}%
\bibitem [{\citenamefont {Bergeret}\ and\ \citenamefont
  {Tokatly}(2013)}]{Bergeret_PRL2013}%
  \BibitemOpen
  \bibfield  {author} {\bibinfo {author} {\bibfnamefont {F.~S.}\ \bibnamefont
  {Bergeret}}\ and\ \bibinfo {author} {\bibfnamefont {I.~V.}\ \bibnamefont
  {Tokatly}},\ }\href {\doibase 10.1103/PhysRevLett.110.117003} {\bibfield
  {journal} {\bibinfo  {journal} {Phys. Rev. Lett.}\ }\textbf {\bibinfo
  {volume} {110}},\ \bibinfo {pages} {117003} (\bibinfo {year}
  {2013})}\BibitemShut {NoStop}%
\bibitem [{\citenamefont {Bergeret}\ and\ \citenamefont
  {Tokatly}(2014)}]{Bergeret_PRB2014}%
  \BibitemOpen
  \bibfield  {author} {\bibinfo {author} {\bibfnamefont {F.~S.}\ \bibnamefont
  {Bergeret}}\ and\ \bibinfo {author} {\bibfnamefont {I.~V.}\ \bibnamefont
  {Tokatly}},\ }\href {\doibase 10.1103/PhysRevB.89.134517} {\bibfield
  {journal} {\bibinfo  {journal} {Phys. Rev. B}\ }\textbf {\bibinfo {volume}
  {89}},\ \bibinfo {pages} {134517} (\bibinfo {year} {2014})}\BibitemShut
  {NoStop}%
\bibitem [{\citenamefont {Jacobsen}\ \emph {et~al.}(2015)\citenamefont
  {Jacobsen}, \citenamefont {Ouassou},\ and\ \citenamefont
  {Linder}}]{Linder_PRB2015}%
  \BibitemOpen
  \bibfield  {author} {\bibinfo {author} {\bibfnamefont {S.~H.}\ \bibnamefont
  {Jacobsen}}, \bibinfo {author} {\bibfnamefont {J.~A.}\ \bibnamefont
  {Ouassou}}, \ and\ \bibinfo {author} {\bibfnamefont {J.}~\bibnamefont
  {Linder}},\ }\href {\doibase 10.1103/PhysRevB.92.024510} {\bibfield
  {journal} {\bibinfo  {journal} {Phys. Rev. B}\ }\textbf {\bibinfo {volume}
  {92}},\ \bibinfo {pages} {024510} (\bibinfo {year} {2015})}\BibitemShut
  {NoStop}%
\bibitem [{\citenamefont {H\"ogl}\ \emph {et~al.}(2015)\citenamefont {H\"ogl},
  \citenamefont {Matos-Abiague}, \citenamefont {Zutic},\ and\ \citenamefont
  {Fabian}}]{Zutic_PRL2015}%
  \BibitemOpen
  \bibfield  {author} {\bibinfo {author} {\bibfnamefont {P.}~\bibnamefont
  {H\"ogl}}, \bibinfo {author} {\bibfnamefont {A.}~\bibnamefont
  {Matos-Abiague}}, \bibinfo {author} {\bibfnamefont {I.}~\bibnamefont
  {Zutic}}, \ and\ \bibinfo {author} {\bibfnamefont {J.}~\bibnamefont
  {Fabian}},\ }\href {\doibase 10.1103/PhysRevLett.115.116601} {\bibfield
  {journal} {\bibinfo  {journal} {Phys. Rev. Lett.}\ }\textbf {\bibinfo
  {volume} {115}},\ \bibinfo {pages} {116601} (\bibinfo {year}
  {2015})}\BibitemShut {NoStop}%
\bibitem [{\citenamefont {Grein}\ \emph {et~al.}(2010)\citenamefont {Grein},
  \citenamefont {L\"ofwander}, \citenamefont {Metalidis},\ and\ \citenamefont
  {Eschrig}}]{Grein_PRB2010}%
  \BibitemOpen
  \bibfield  {author} {\bibinfo {author} {\bibfnamefont {R.}~\bibnamefont
  {Grein}}, \bibinfo {author} {\bibfnamefont {T.}~\bibnamefont {L\"ofwander}},
  \bibinfo {author} {\bibfnamefont {G.}~\bibnamefont {Metalidis}}, \ and\
  \bibinfo {author} {\bibfnamefont {M.}~\bibnamefont {Eschrig}},\ }\href
  {\doibase 10.1103/PhysRevB.81.094508} {\bibfield  {journal} {\bibinfo
  {journal} {Phys. Rev. B}\ }\textbf {\bibinfo {volume} {81}},\ \bibinfo
  {pages} {094508} (\bibinfo {year} {2010})}\BibitemShut {NoStop}%
\bibitem [{\citenamefont {H\"ubler}\ \emph {et~al.}(2010)\citenamefont
  {H\"ubler}, \citenamefont {Lemyre}, \citenamefont {Beckmann},\ and\
  \citenamefont {v.~L\"ohneysen}}]{Hubler_PRB2010}%
  \BibitemOpen
  \bibfield  {author} {\bibinfo {author} {\bibfnamefont {F.}~\bibnamefont
  {H\"ubler}}, \bibinfo {author} {\bibfnamefont {J.~C.}\ \bibnamefont
  {Lemyre}}, \bibinfo {author} {\bibfnamefont {D.}~\bibnamefont {Beckmann}}, \
  and\ \bibinfo {author} {\bibfnamefont {H.}~\bibnamefont {v.~L\"ohneysen}},\
  }\href {\doibase 10.1103/PhysRevB.81.184524} {\bibfield  {journal} {\bibinfo
  {journal} {Phys. Rev. B}\ }\textbf {\bibinfo {volume} {81}},\ \bibinfo
  {pages} {184524} (\bibinfo {year} {2010})}\BibitemShut {NoStop}%
\bibitem [{\citenamefont {H\"ubler}\ \emph {et~al.}(2012)\citenamefont
  {H\"ubler}, \citenamefont {Wolf}, \citenamefont {Beckmann},\ and\
  \citenamefont {v.~L\"ohneysen}}]{Hubler_PRL2012}%
  \BibitemOpen
  \bibfield  {author} {\bibinfo {author} {\bibfnamefont {F.}~\bibnamefont
  {H\"ubler}}, \bibinfo {author} {\bibfnamefont {M.~J.}\ \bibnamefont {Wolf}},
  \bibinfo {author} {\bibfnamefont {D.}~\bibnamefont {Beckmann}}, \ and\
  \bibinfo {author} {\bibfnamefont {H.}~\bibnamefont {v.~L\"ohneysen}},\ }\href
  {\doibase 10.1103/PhysRevLett.109.207001} {\bibfield  {journal} {\bibinfo
  {journal} {Phys. Rev. Lett.}\ }\textbf {\bibinfo {volume} {109}},\ \bibinfo
  {pages} {207001} (\bibinfo {year} {2012})}\BibitemShut {NoStop}%
\bibitem [{\citenamefont {Wolf}\ \emph {et~al.}(2013)\citenamefont {Wolf},
  \citenamefont {H\"ubler}, \citenamefont {Kolenda}, \citenamefont
  {v.~L\"ohneysen},\ and\ \citenamefont {Beckmann}}]{Hubler_PRB2013}%
  \BibitemOpen
  \bibfield  {author} {\bibinfo {author} {\bibfnamefont {M.~J.}\ \bibnamefont
  {Wolf}}, \bibinfo {author} {\bibfnamefont {F.}~\bibnamefont {H\"ubler}},
  \bibinfo {author} {\bibfnamefont {S.}~\bibnamefont {Kolenda}}, \bibinfo
  {author} {\bibfnamefont {H.}~\bibnamefont {v.~L\"ohneysen}}, \ and\ \bibinfo
  {author} {\bibfnamefont {D.}~\bibnamefont {Beckmann}},\ }\href {\doibase
  10.1103/PhysRevB.87.024517} {\bibfield  {journal} {\bibinfo  {journal} {Phys.
  Rev. B}\ }\textbf {\bibinfo {volume} {87}},\ \bibinfo {pages} {024517}
  (\bibinfo {year} {2013})}\BibitemShut {NoStop}%
\bibitem [{\citenamefont {Quay}\ \emph {et~al.}(2013)\citenamefont {Quay},
  \citenamefont {Chevallier}, \citenamefont {Bena},\ and\ \citenamefont
  {Aprili}}]{quay_NatPhys2013}%
  \BibitemOpen
  \bibfield  {author} {\bibinfo {author} {\bibfnamefont {C.}~\bibnamefont
  {Quay}}, \bibinfo {author} {\bibfnamefont {D.}~\bibnamefont {Chevallier}},
  \bibinfo {author} {\bibfnamefont {C.}~\bibnamefont {Bena}}, \ and\ \bibinfo
  {author} {\bibfnamefont {M.}~\bibnamefont {Aprili}},\ }\href {\doibase
  10.1038/nphys2518} {\bibfield  {journal} {\bibinfo  {journal} {Nat. Phys.}\
  }\textbf {\bibinfo {volume} {9}},\ \bibinfo {pages} {84} (\bibinfo {year}
  {2013})}\BibitemShut {NoStop}%
\bibitem [{\citenamefont {Wakamura}\ \emph {et~al.}(2014)\citenamefont
  {Wakamura}, \citenamefont {Hasegawa}, \citenamefont {Ohnishi}, \citenamefont
  {Niimi},\ and\ \citenamefont {Otani}}]{Wakamura_PRL2014}%
  \BibitemOpen
  \bibfield  {author} {\bibinfo {author} {\bibfnamefont {T.}~\bibnamefont
  {Wakamura}}, \bibinfo {author} {\bibfnamefont {N.}~\bibnamefont {Hasegawa}},
  \bibinfo {author} {\bibfnamefont {K.}~\bibnamefont {Ohnishi}}, \bibinfo
  {author} {\bibfnamefont {Y.}~\bibnamefont {Niimi}}, \ and\ \bibinfo {author}
  {\bibfnamefont {Y.}~\bibnamefont {Otani}},\ }\href {\doibase
  10.1103/PhysRevLett.112.036602} {\bibfield  {journal} {\bibinfo  {journal}
  {Phys. Rev. Lett.}\ }\textbf {\bibinfo {volume} {112}},\ \bibinfo {pages}
  {036602} (\bibinfo {year} {2014})}\BibitemShut {NoStop}%
\bibitem [{\citenamefont {Beckmann}(2016)}]{Beckmann_JPhysConMat2016}%
  \BibitemOpen
  \bibfield  {author} {\bibinfo {author} {\bibfnamefont {D.}~\bibnamefont
  {Beckmann}},\ }\href {http://stacks.iop.org/0953-8984/28/i=16/a=163001}
  {\bibfield  {journal} {\bibinfo  {journal} {J. Phys.: Condens. Matter}\
  }\textbf {\bibinfo {volume} {28}},\ \bibinfo {pages} {163001} (\bibinfo
  {year} {2016})}\BibitemShut {NoStop}%
\bibitem [{\citenamefont {Poli}\ \emph {et~al.}(2008)\citenamefont {Poli},
  \citenamefont {Morten}, \citenamefont {Urech}, \citenamefont {Brataas},
  \citenamefont {Haviland},\ and\ \citenamefont {Korenivski}}]{Poli_PRL2008}%
  \BibitemOpen
  \bibfield  {author} {\bibinfo {author} {\bibfnamefont {N.}~\bibnamefont
  {Poli}}, \bibinfo {author} {\bibfnamefont {J.~P.}\ \bibnamefont {Morten}},
  \bibinfo {author} {\bibfnamefont {M.}~\bibnamefont {Urech}}, \bibinfo
  {author} {\bibfnamefont {A.}~\bibnamefont {Brataas}}, \bibinfo {author}
  {\bibfnamefont {D.~B.}\ \bibnamefont {Haviland}}, \ and\ \bibinfo {author}
  {\bibfnamefont {V.}~\bibnamefont {Korenivski}},\ }\href {\doibase
  10.1103/PhysRevLett.100.136601} {\bibfield  {journal} {\bibinfo  {journal}
  {Phys. Rev. Lett.}\ }\textbf {\bibinfo {volume} {100}},\ \bibinfo {pages}
  {136601} (\bibinfo {year} {2008})}\BibitemShut {NoStop}%
\bibitem [{\citenamefont {Yang}\ \emph {et~al.}(2010)\citenamefont {Yang},
  \citenamefont {Yang}, \citenamefont {Takahashi}, \citenamefont {Maekawa},\
  and\ \citenamefont {Parkin}}]{yang_NatMat2010}%
  \BibitemOpen
  \bibfield  {author} {\bibinfo {author} {\bibfnamefont {H.}~\bibnamefont
  {Yang}}, \bibinfo {author} {\bibfnamefont {S.-H.}\ \bibnamefont {Yang}},
  \bibinfo {author} {\bibfnamefont {S.}~\bibnamefont {Takahashi}}, \bibinfo
  {author} {\bibfnamefont {S.}~\bibnamefont {Maekawa}}, \ and\ \bibinfo
  {author} {\bibfnamefont {S.~S.}\ \bibnamefont {Parkin}},\ }\href {\doibase
  10.1038/nmat2781} {\bibfield  {journal} {\bibinfo  {journal} {Nat. Mater.}\
  }\textbf {\bibinfo {volume} {9}},\ \bibinfo {pages} {586} (\bibinfo {year}
  {2010})}\BibitemShut {NoStop}%
\bibitem [{\citenamefont {Wakamura}\ \emph {et~al.}(2015)\citenamefont
  {Wakamura}, \citenamefont {Akaike}, \citenamefont {Omori}, \citenamefont
  {Niimi}, \citenamefont {Takahashi}, \citenamefont {Fujimaki}, \citenamefont
  {Maekawa},\ and\ \citenamefont {Otani}}]{wakamura_NatMat2015}%
  \BibitemOpen
  \bibfield  {author} {\bibinfo {author} {\bibfnamefont {T.}~\bibnamefont
  {Wakamura}}, \bibinfo {author} {\bibfnamefont {H.}~\bibnamefont {Akaike}},
  \bibinfo {author} {\bibfnamefont {Y.}~\bibnamefont {Omori}}, \bibinfo
  {author} {\bibfnamefont {Y.}~\bibnamefont {Niimi}}, \bibinfo {author}
  {\bibfnamefont {S.}~\bibnamefont {Takahashi}}, \bibinfo {author}
  {\bibfnamefont {A.}~\bibnamefont {Fujimaki}}, \bibinfo {author}
  {\bibfnamefont {S.}~\bibnamefont {Maekawa}}, \ and\ \bibinfo {author}
  {\bibfnamefont {Y.}~\bibnamefont {Otani}},\ }\href {\doibase
  10.1038/nmat4276} {\bibfield  {journal} {\bibinfo  {journal} {Nat. Mater.}\
  }\textbf {\bibinfo {volume} {14}},\ \bibinfo {pages} {675} (\bibinfo {year}
  {2015})}\BibitemShut {NoStop}%
\bibitem [{\citenamefont {Inoue}\ \emph {et~al.}(2017)\citenamefont {Inoue},
  \citenamefont {Ichioka},\ and\ \citenamefont {Adachi}}]{Inoue_PRB2017}%
  \BibitemOpen
  \bibfield  {author} {\bibinfo {author} {\bibfnamefont {M.}~\bibnamefont
  {Inoue}}, \bibinfo {author} {\bibfnamefont {M.}~\bibnamefont {Ichioka}}, \
  and\ \bibinfo {author} {\bibfnamefont {H.}~\bibnamefont {Adachi}},\ }\href
  {\doibase 10.1103/PhysRevB.96.024414} {\bibfield  {journal} {\bibinfo
  {journal} {Phys. Rev. B}\ }\textbf {\bibinfo {volume} {96}},\ \bibinfo
  {pages} {024414} (\bibinfo {year} {2017})}\BibitemShut {NoStop}%
\bibitem [{\citenamefont {Bell}\ \emph {et~al.}(2008)\citenamefont {Bell},
  \citenamefont {Milikisyants}, \citenamefont {Huber},\ and\ \citenamefont
  {Aarts}}]{Bell_PRL2008}%
  \BibitemOpen
  \bibfield  {author} {\bibinfo {author} {\bibfnamefont {C.}~\bibnamefont
  {Bell}}, \bibinfo {author} {\bibfnamefont {S.}~\bibnamefont {Milikisyants}},
  \bibinfo {author} {\bibfnamefont {M.}~\bibnamefont {Huber}}, \ and\ \bibinfo
  {author} {\bibfnamefont {J.}~\bibnamefont {Aarts}},\ }\href {\doibase
  10.1103/PhysRevLett.100.047002} {\bibfield  {journal} {\bibinfo  {journal}
  {Phys. Rev. Lett.}\ }\textbf {\bibinfo {volume} {100}},\ \bibinfo {pages}
  {047002} (\bibinfo {year} {2008})}\BibitemShut {NoStop}%
\bibitem [{\citenamefont {Jeon}\ \emph {et~al.}(2018)\citenamefont {Jeon},
  \citenamefont {Ciccarelli}, \citenamefont {Ferguson}, \citenamefont
  {Kurebayashi}, \citenamefont {Cohen}, \citenamefont {Montiel}, \citenamefont
  {Eschrig}, \citenamefont {Robinson},\ and\ \citenamefont
  {Blamire}}]{Jeon_NatMat2018}%
  \BibitemOpen
  \bibfield  {author} {\bibinfo {author} {\bibfnamefont {K.-R.}\ \bibnamefont
  {Jeon}}, \bibinfo {author} {\bibfnamefont {C.}~\bibnamefont {Ciccarelli}},
  \bibinfo {author} {\bibfnamefont {A.~J.}\ \bibnamefont {Ferguson}}, \bibinfo
  {author} {\bibfnamefont {H.}~\bibnamefont {Kurebayashi}}, \bibinfo {author}
  {\bibfnamefont {L.~F.}\ \bibnamefont {Cohen}}, \bibinfo {author}
  {\bibfnamefont {X.}~\bibnamefont {Montiel}}, \bibinfo {author} {\bibfnamefont
  {M.}~\bibnamefont {Eschrig}}, \bibinfo {author} {\bibfnamefont {J.~W.~A.}\
  \bibnamefont {Robinson}}, \ and\ \bibinfo {author} {\bibfnamefont {M.~G.}\
  \bibnamefont {Blamire}},\ }\href {\doibase 10.1038/s41563-018-0058-9}
  {\bibfield  {journal} {\bibinfo  {journal} {Nat. Mater.}\ }\textbf {\bibinfo
  {volume} {17}},\ \bibinfo {pages} {499} (\bibinfo {year} {2018})}\BibitemShut
  {NoStop}%
\bibitem [{\citenamefont {Tserkovnyak}\ \emph {et~al.}(2005)\citenamefont
  {Tserkovnyak}, \citenamefont {Brataas}, \citenamefont {Bauer},\ and\
  \citenamefont {Halperin}}]{Tserkovnyak_RevModPhys2005}%
  \BibitemOpen
  \bibfield  {author} {\bibinfo {author} {\bibfnamefont {Y.}~\bibnamefont
  {Tserkovnyak}}, \bibinfo {author} {\bibfnamefont {A.}~\bibnamefont
  {Brataas}}, \bibinfo {author} {\bibfnamefont {G.~E.~W.}\ \bibnamefont
  {Bauer}}, \ and\ \bibinfo {author} {\bibfnamefont {B.~I.}\ \bibnamefont
  {Halperin}},\ }\href {\doibase 10.1103/RevModPhys.77.1375} {\bibfield
  {journal} {\bibinfo  {journal} {Rev. Mod. Phys.}\ }\textbf {\bibinfo {volume}
  {77}},\ \bibinfo {pages} {1375} (\bibinfo {year} {2005})}\BibitemShut
  {NoStop}%
\bibitem [{\citenamefont {Morten}\ \emph {et~al.}(2008)\citenamefont {Morten},
  \citenamefont {Brataas}, \citenamefont {Bauer}, \citenamefont {Belzig},\ and\
  \citenamefont {Tserkovnyak}}]{Morten_EPL2008}%
  \BibitemOpen
  \bibfield  {author} {\bibinfo {author} {\bibfnamefont {J.~P.}\ \bibnamefont
  {Morten}}, \bibinfo {author} {\bibfnamefont {A.}~\bibnamefont {Brataas}},
  \bibinfo {author} {\bibfnamefont {G.~E.~W.}\ \bibnamefont {Bauer}}, \bibinfo
  {author} {\bibfnamefont {W.}~\bibnamefont {Belzig}}, \ and\ \bibinfo {author}
  {\bibfnamefont {Y.}~\bibnamefont {Tserkovnyak}},\ }\href
  {http://stacks.iop.org/0295-5075/84/i=5/a=57008} {\bibfield  {journal}
  {\bibinfo  {journal} {Europhys. Lett.}\ }\textbf {\bibinfo {volume} {84}},\
  \bibinfo {pages} {57008} (\bibinfo {year} {2008})}\BibitemShut {NoStop}%
\bibitem [{\citenamefont {Montiel}\ and\ \citenamefont
  {Eschrig}(2018)}]{Montiel_PRB2018}%
  \BibitemOpen
  \bibfield  {author} {\bibinfo {author} {\bibfnamefont {X.}~\bibnamefont
  {Montiel}}\ and\ \bibinfo {author} {\bibfnamefont {M.}~\bibnamefont
  {Eschrig}},\ }\href {\doibase 10.1103/PhysRevB.98.104513} {\bibfield
  {journal} {\bibinfo  {journal} {Phys. Rev. B}\ }\textbf {\bibinfo {volume}
  {98}},\ \bibinfo {pages} {104513} (\bibinfo {year} {2018})}\BibitemShut
  {NoStop}%
\bibitem [{\citenamefont {Kim}\ \emph {et~al.}(2018)\citenamefont {Kim},
  \citenamefont {Myers},\ and\ \citenamefont
  {Tserkovnyak}}]{Tserkovnyak_PRL_2018}%
  \BibitemOpen
  \bibfield  {author} {\bibinfo {author} {\bibfnamefont {S.~K.}\ \bibnamefont
  {Kim}}, \bibinfo {author} {\bibfnamefont {R.}~\bibnamefont {Myers}}, \ and\
  \bibinfo {author} {\bibfnamefont {Y.}~\bibnamefont {Tserkovnyak}},\ }\href
  {\doibase 10.1103/PhysRevLett.121.187203} {\bibfield  {journal} {\bibinfo
  {journal} {Phys. Rev. Lett.}\ }\textbf {\bibinfo {volume} {121}},\ \bibinfo
  {pages} {187203} (\bibinfo {year} {2018})}\BibitemShut {NoStop}%
\bibitem [{\citenamefont {Taira}\ \emph {et~al.}(2018)\citenamefont {Taira},
  \citenamefont {Ichioka}, \citenamefont {Takei},\ and\ \citenamefont
  {Adachi}}]{Taira_PRB_2018}%
  \BibitemOpen
  \bibfield  {author} {\bibinfo {author} {\bibfnamefont {T.}~\bibnamefont
  {Taira}}, \bibinfo {author} {\bibfnamefont {M.}~\bibnamefont {Ichioka}},
  \bibinfo {author} {\bibfnamefont {S.}~\bibnamefont {Takei}}, \ and\ \bibinfo
  {author} {\bibfnamefont {H.}~\bibnamefont {Adachi}},\ }\href {\doibase
  10.1103/PhysRevB.98.214437} {\bibfield  {journal} {\bibinfo  {journal} {Phys.
  Rev. B}\ }\textbf {\bibinfo {volume} {98}},\ \bibinfo {pages} {214437}
  (\bibinfo {year} {2018})}\BibitemShut {NoStop}%
\bibitem [{\citenamefont {{Jeon}}\ \emph {et~al.}(2019)\citenamefont {{Jeon}},
  \citenamefont {{Montiel}}, \citenamefont {{Komori}}, \citenamefont
  {{Ciccarelli}}, \citenamefont {{Haigh}}, \citenamefont {{Kurebayashi}},
  \citenamefont {{Cohen}}, \citenamefont {{Lee}}, \citenamefont {{Blamire}},\
  and\ \citenamefont {{Robinson}}}]{Jeon_PRX_2020}%
  \BibitemOpen
  \bibfield  {author} {\bibinfo {author} {\bibfnamefont {K.~R.}\ \bibnamefont
  {{Jeon}}}, \bibinfo {author} {\bibfnamefont {X.}~\bibnamefont {{Montiel}}},
  \bibinfo {author} {\bibfnamefont {S.}~\bibnamefont {{Komori}}}, \bibinfo
  {author} {\bibfnamefont {C.}~\bibnamefont {{Ciccarelli}}}, \bibinfo {author}
  {\bibfnamefont {J.}~\bibnamefont {{Haigh}}}, \bibinfo {author} {\bibfnamefont
  {H.}~\bibnamefont {{Kurebayashi}}}, \bibinfo {author} {\bibfnamefont {L.~F.}\
  \bibnamefont {{Cohen}}}, \bibinfo {author} {\bibfnamefont {C.~M.}\
  \bibnamefont {{Lee}}}, \bibinfo {author} {\bibfnamefont {M.~G.}\ \bibnamefont
  {{Blamire}}}, \ and\ \bibinfo {author} {\bibfnamefont {J.~W.~A.}\
  \bibnamefont {{Robinson}}},\ }\href {https://arxiv.org/abs/1908.00873}
  {\bibfield  {journal} {\bibinfo  {journal} {arXiv:1908.00873}\ } (\bibinfo
  {year} {2019})}\BibitemShut {NoStop}%
\bibitem [{\citenamefont {Silaev}(2020)}]{Silaev_PRB2020}%
  \BibitemOpen
  \bibfield  {author} {\bibinfo {author} {\bibfnamefont {M.~A.}\ \bibnamefont
  {Silaev}},\ }\href {\doibase 10.1103/PhysRevB.102.180502} {\bibfield
  {journal} {\bibinfo  {journal} {Phys. Rev. B}\ }\textbf {\bibinfo {volume}
  {102}},\ \bibinfo {pages} {180502(R)} (\bibinfo {year} {2020})}\BibitemShut
  {NoStop}%
\bibitem [{\citenamefont {Simensen}\ \emph {et~al.}(2021)\citenamefont
  {Simensen}, \citenamefont {Johnsen}, \citenamefont {Linder},\ and\
  \citenamefont {Brataas}}]{Simensen_PRB2021}%
  \BibitemOpen
  \bibfield  {author} {\bibinfo {author} {\bibfnamefont {H.~T.}\ \bibnamefont
  {Simensen}}, \bibinfo {author} {\bibfnamefont {L.~G.}\ \bibnamefont
  {Johnsen}}, \bibinfo {author} {\bibfnamefont {J.}~\bibnamefont {Linder}}, \
  and\ \bibinfo {author} {\bibfnamefont {A.}~\bibnamefont {Brataas}},\ }\href
  {\doibase 10.1103/PhysRevB.103.024524} {\bibfield  {journal} {\bibinfo
  {journal} {Phys. Rev. B}\ }\textbf {\bibinfo {volume} {103}},\ \bibinfo
  {pages} {024524} (\bibinfo {year} {2021})}\BibitemShut {NoStop}%
\bibitem [{\citenamefont {Chan}\ \emph {et~al.}(0221)\citenamefont {Chan},
  \citenamefont {Cubukcu}, \citenamefont {Montiel}, \citenamefont {Komori},
  \citenamefont {Vanstone}, \citenamefont {Thompson}, \citenamefont {Perkins},
  \citenamefont {Blamire}, \citenamefont {Robinson}, \citenamefont {Eschrig},
  \citenamefont {Kurebayashi},\ and\ \citenamefont {Cohen}}]{Chan_arxiv2022}%
  \BibitemOpen
  \bibfield  {author} {\bibinfo {author} {\bibfnamefont {A.~K.}\ \bibnamefont
  {Chan}}, \bibinfo {author} {\bibfnamefont {M.}~\bibnamefont {Cubukcu}},
  \bibinfo {author} {\bibfnamefont {X.}~\bibnamefont {Montiel}}, \bibinfo
  {author} {\bibfnamefont {S.}~\bibnamefont {Komori}}, \bibinfo {author}
  {\bibfnamefont {A.}~\bibnamefont {Vanstone}}, \bibinfo {author}
  {\bibfnamefont {J.~E.}\ \bibnamefont {Thompson}}, \bibinfo {author}
  {\bibfnamefont {G.~K.}\ \bibnamefont {Perkins}}, \bibinfo {author}
  {\bibfnamefont {M.}~\bibnamefont {Blamire}}, \bibinfo {author} {\bibfnamefont
  {J.}~\bibnamefont {Robinson}}, \bibinfo {author} {\bibfnamefont
  {M.}~\bibnamefont {Eschrig}}, \bibinfo {author} {\bibfnamefont
  {H.}~\bibnamefont {Kurebayashi}}, \ and\ \bibinfo {author} {\bibfnamefont
  {L.~F.}\ \bibnamefont {Cohen}},\ }\href
  {https://doi.org/10.48550/arXiv.2202.01520} {\bibfield  {journal} {\bibinfo
  {journal} {arXiv:2202.01520}\ } (\bibinfo {year} {20221})}\BibitemShut
  {NoStop}%
\bibitem [{\citenamefont {Houzet}(2008)}]{Houzet_PRL101_2008}%
  \BibitemOpen
  \bibfield  {author} {\bibinfo {author} {\bibfnamefont {M.}~\bibnamefont
  {Houzet}},\ }\href {\doibase 10.1103/PhysRevLett.101.057009} {\bibfield
  {journal} {\bibinfo  {journal} {Phys. Rev. Lett.}\ }\textbf {\bibinfo
  {volume} {101}},\ \bibinfo {pages} {057009} (\bibinfo {year}
  {2008})}\BibitemShut {NoStop}%
\bibitem [{\citenamefont {Usadel}(1970)}]{Usadel_PRL1970}%
  \BibitemOpen
  \bibfield  {author} {\bibinfo {author} {\bibfnamefont {K.~D.}\ \bibnamefont
  {Usadel}},\ }\href {\doibase 10.1103/PhysRevLett.25.507} {\bibfield
  {journal} {\bibinfo  {journal} {Phys. Rev. Lett.}\ }\textbf {\bibinfo
  {volume} {25}},\ \bibinfo {pages} {507} (\bibinfo {year} {1970})}\BibitemShut
  {NoStop}%
\bibitem [{\citenamefont {Belzig}\ \emph {et~al.}(1999)\citenamefont {Belzig},
  \citenamefont {Wilhelm}, \citenamefont {Bruder}, \citenamefont {Schön},\
  and\ \citenamefont {Zaikin}}]{Belzig_SuperLattice1999}%
  \BibitemOpen
  \bibfield  {author} {\bibinfo {author} {\bibfnamefont {W.}~\bibnamefont
  {Belzig}}, \bibinfo {author} {\bibfnamefont {F.~K.}\ \bibnamefont {Wilhelm}},
  \bibinfo {author} {\bibfnamefont {C.}~\bibnamefont {Bruder}}, \bibinfo
  {author} {\bibfnamefont {G.}~\bibnamefont {Schön}}, \ and\ \bibinfo {author}
  {\bibfnamefont {A.~D.}\ \bibnamefont {Zaikin}},\ }\href {\doibase
  https://doi.org/10.1006/spmi.1999.0710} {\bibfield  {journal} {\bibinfo
  {journal} {Superlattices Microstruc.}\ }\textbf {\bibinfo {volume} {25}},\
  \bibinfo {pages} {1251 } (\bibinfo {year} {1999})}\BibitemShut {NoStop}%
\bibitem [{\citenamefont {Eschrig}(2009)}]{Eschrig_PRB2009}%
  \BibitemOpen
  \bibfield  {author} {\bibinfo {author} {\bibfnamefont {M.}~\bibnamefont
  {Eschrig}},\ }\href {\doibase 10.1103/PhysRevB.80.134511} {\bibfield
  {journal} {\bibinfo  {journal} {Phys. Rev. B}\ }\textbf {\bibinfo {volume}
  {80}},\ \bibinfo {pages} {134511} (\bibinfo {year} {2009})}\BibitemShut
  {NoStop}%
\bibitem [{\citenamefont {Eilenberger}(1968)}]{Eilenberger_1968}%
  \BibitemOpen
  \bibfield  {author} {\bibinfo {author} {\bibfnamefont {G.}~\bibnamefont
  {Eilenberger}},\ }\href {\doibase 10.1007/BF01379803} {\bibfield  {journal}
  {\bibinfo  {journal} {Z. Phys.}\ }\textbf {\bibinfo {volume} {214}},\
  \bibinfo {pages} {195} (\bibinfo {year} {1968})}\BibitemShut {NoStop}%
\bibitem [{\citenamefont {Larkin}\ and\ \citenamefont
  {Ovchinnikov}(1969)}]{larkin_JETP1969}%
  \BibitemOpen
  \bibfield  {author} {\bibinfo {author} {\bibfnamefont {A.}~\bibnamefont
  {Larkin}}\ and\ \bibinfo {author} {\bibfnamefont {Y.~N.}\ \bibnamefont
  {Ovchinnikov}},\ }\href
  {http://www.jetp.ac.ru/cgi-bin/e/index/e/28/6/p1200?a=list} {\bibfield
  {journal} {\bibinfo  {journal} {Sov Phys JETP}\ }\textbf {\bibinfo {volume}
  {28}},\ \bibinfo {pages} {1200} (\bibinfo {year} {1969})}\BibitemShut
  {NoStop}%
\bibitem [{\citenamefont {Gor'kov}\ and\ \citenamefont
  {Rusinov}(1964)}]{Rusinov_JETP1964}%
  \BibitemOpen
  \bibfield  {author} {\bibinfo {author} {\bibfnamefont {L.}~\bibnamefont
  {Gor'kov}}\ and\ \bibinfo {author} {\bibfnamefont {A.}~\bibnamefont
  {Rusinov}},\ }\href
  {http://www.jetp.ac.ru/cgi-bin/e/index/e/19/4/p922?a=list} {\bibfield
  {journal} {\bibinfo  {journal} {Sov. Phys.--JETP 19, 922.[Zh. Eksp. Teor.
  Fiz. 46, 1363.]}\ } (\bibinfo {year} {1964})}\BibitemShut {NoStop}%
\bibitem [{\citenamefont {Kuprianov}\ and\ \citenamefont
  {Lukichev}(1988)}]{Kuprianov_JETP1988}%
  \BibitemOpen
  \bibfield  {author} {\bibinfo {author} {\bibfnamefont {M.~Y.}\ \bibnamefont
  {Kuprianov}}\ and\ \bibinfo {author} {\bibfnamefont {V.~F.}\ \bibnamefont
  {Lukichev}},\ }\href
  {http://www.jetp.ac.ru/cgi-bin/e/index/e/67/6/p1163?a=list} {\bibfield
  {journal} {\bibinfo  {journal} {Sov. Phys.--JETP 67,1163 [Zh. Eksp. Teor.
  Fiz. 94, 139.]}\ } (\bibinfo {year} {1988})}\BibitemShut {NoStop}%
\bibitem [{\citenamefont {Nazarov}(1994)}]{Nazarov_PRL1994}%
  \BibitemOpen
  \bibfield  {author} {\bibinfo {author} {\bibfnamefont {Y.~V.}\ \bibnamefont
  {Nazarov}},\ }\href {\doibase 10.1103/PhysRevLett.73.1420} {\bibfield
  {journal} {\bibinfo  {journal} {Phys. Rev. Lett.}\ }\textbf {\bibinfo
  {volume} {73}},\ \bibinfo {pages} {1420} (\bibinfo {year}
  {1994})}\BibitemShut {NoStop}%
\bibitem [{\citenamefont {Nazarov}(1999)}]{Nazarov_SuperLattMicro1999}%
  \BibitemOpen
  \bibfield  {author} {\bibinfo {author} {\bibfnamefont {Y.~V.}\ \bibnamefont
  {Nazarov}},\ }\href {\doibase https://doi.org/10.1006/spmi.1999.0738}
  {\bibfield  {journal} {\bibinfo  {journal} {Superlattices Microstruc.}\
  }\textbf {\bibinfo {volume} {25}},\ \bibinfo {pages} {1221 } (\bibinfo {year}
  {1999})}\BibitemShut {NoStop}%
\bibitem [{\citenamefont {Eschrig}\ \emph {et~al.}(2015)\citenamefont
  {Eschrig}, \citenamefont {Cottet}, \citenamefont {Belzig},\ and\
  \citenamefont {Linder}}]{Eschrig_NJPhys2015}%
  \BibitemOpen
  \bibfield  {author} {\bibinfo {author} {\bibfnamefont {M.}~\bibnamefont
  {Eschrig}}, \bibinfo {author} {\bibfnamefont {A.}~\bibnamefont {Cottet}},
  \bibinfo {author} {\bibfnamefont {W.}~\bibnamefont {Belzig}}, \ and\ \bibinfo
  {author} {\bibfnamefont {J.}~\bibnamefont {Linder}},\ }\href
  {http://stacks.iop.org/1367-2630/17/i=8/a=083037} {\bibfield  {journal}
  {\bibinfo  {journal} {New J. Phys.}\ }\textbf {\bibinfo {volume} {17}},\
  \bibinfo {pages} {083037} (\bibinfo {year} {2015})}\BibitemShut {NoStop}%
\bibitem [{\citenamefont {Jacobsen}\ \emph {et~al.}(2016)\citenamefont
  {Jacobsen}, \citenamefont {Kulagina},\ and\ \citenamefont
  {Linder}}]{Jacobsen_SciRep2016}%
  \BibitemOpen
  \bibfield  {author} {\bibinfo {author} {\bibfnamefont {S.~H.}\ \bibnamefont
  {Jacobsen}}, \bibinfo {author} {\bibfnamefont {I.}~\bibnamefont {Kulagina}},
  \ and\ \bibinfo {author} {\bibfnamefont {J.}~\bibnamefont {Linder}},\ }\href
  {https://www.nature.com/articles/srep23926} {\bibfield  {journal} {\bibinfo
  {journal} {Sci. Rep.}\ }\textbf {\bibinfo {volume} {6}},\ \bibinfo {pages}
  {23926} (\bibinfo {year} {2016})}\BibitemShut {NoStop}%
\bibitem [{\citenamefont {Ouassou}\ \emph {et~al.}(2017)\citenamefont
  {Ouassou}, \citenamefont {Jacobsen},\ and\ \citenamefont
  {Linder}}]{Jacobsen_PRB2017}%
  \BibitemOpen
  \bibfield  {author} {\bibinfo {author} {\bibfnamefont {J.~A.}\ \bibnamefont
  {Ouassou}}, \bibinfo {author} {\bibfnamefont {S.~H.}\ \bibnamefont
  {Jacobsen}}, \ and\ \bibinfo {author} {\bibfnamefont {J.}~\bibnamefont
  {Linder}},\ }\href {\doibase 10.1103/PhysRevB.96.094505} {\bibfield
  {journal} {\bibinfo  {journal} {Phys. Rev. B}\ }\textbf {\bibinfo {volume}
  {96}},\ \bibinfo {pages} {094505} (\bibinfo {year} {2017})}\BibitemShut
  {NoStop}%
\bibitem [{\citenamefont {Robinson}\ \emph {et~al.}(2014)\citenamefont
  {Robinson}, \citenamefont {Banerjee},\ and\ \citenamefont
  {Blamire}}]{Robinson_PRB2014}%
  \BibitemOpen
  \bibfield  {author} {\bibinfo {author} {\bibfnamefont {J.~W.~A.}\
  \bibnamefont {Robinson}}, \bibinfo {author} {\bibfnamefont {N.}~\bibnamefont
  {Banerjee}}, \ and\ \bibinfo {author} {\bibfnamefont {M.~G.}\ \bibnamefont
  {Blamire}},\ }\href {\doibase 10.1103/PhysRevB.89.104505} {\bibfield
  {journal} {\bibinfo  {journal} {Phys. Rev. B}\ }\textbf {\bibinfo {volume}
  {89}},\ \bibinfo {pages} {104505} (\bibinfo {year} {2014})}\BibitemShut
  {NoStop}%
\bibitem [{\citenamefont {Ishikawa}\ \emph {et~al.}(1965)\citenamefont
  {Ishikawa}, \citenamefont {Tournier},\ and\ \citenamefont
  {Filippi}}]{Ishikawa_JPCS1965}%
  \BibitemOpen
  \bibfield  {author} {\bibinfo {author} {\bibfnamefont {Y.}~\bibnamefont
  {Ishikawa}}, \bibinfo {author} {\bibfnamefont {R.}~\bibnamefont {Tournier}},
  \ and\ \bibinfo {author} {\bibfnamefont {J.}~\bibnamefont {Filippi}},\ }\href
  {\doibase https://doi.org/10.1016/0022-3697(65)90204-0} {\bibfield  {journal}
  {\bibinfo  {journal} {J. Phys. Chem. Solids}\ }\textbf {\bibinfo {volume}
  {26}},\ \bibinfo {pages} {1727 } (\bibinfo {year} {1965})}\BibitemShut
  {NoStop}%
\bibitem [{\citenamefont {Strom-Olsen}\ \emph {et~al.}(1979)\citenamefont
  {Strom-Olsen}, \citenamefont {Wilford}, \citenamefont {Burke},\ and\
  \citenamefont {Rainford}}]{Strom_Olsen_1979}%
  \BibitemOpen
  \bibfield  {author} {\bibinfo {author} {\bibfnamefont {J.~O.}\ \bibnamefont
  {Strom-Olsen}}, \bibinfo {author} {\bibfnamefont {D.~F.}\ \bibnamefont
  {Wilford}}, \bibinfo {author} {\bibfnamefont {S.~K.}\ \bibnamefont {Burke}},
  \ and\ \bibinfo {author} {\bibfnamefont {B.~D.}\ \bibnamefont {Rainford}},\
  }\href {\doibase 10.1088/0305-4608/9/5/002} {\bibfield  {journal} {\bibinfo
  {journal} {J. Phys. F: Met. Phys.}\ }\textbf {\bibinfo {volume} {9}},\
  \bibinfo {pages} {L95} (\bibinfo {year} {1979})}\BibitemShut {NoStop}%
\bibitem [{\citenamefont {Babic}\ \emph {et~al.}(1980)\citenamefont {Babic},
  \citenamefont {Kajzar},\ and\ \citenamefont {Parette}}]{Babic_JPCS1980}%
  \BibitemOpen
  \bibfield  {author} {\bibinfo {author} {\bibfnamefont {B.}~\bibnamefont
  {Babic}}, \bibinfo {author} {\bibfnamefont {F.}~\bibnamefont {Kajzar}}, \
  and\ \bibinfo {author} {\bibfnamefont {G.}~\bibnamefont {Parette}},\ }\href
  {\doibase https://doi.org/10.1016/0022-3697(80)90132-8} {\bibfield  {journal}
  {\bibinfo  {journal} {J. Phys. Chem. Solids}\ }\textbf {\bibinfo {volume}
  {41}},\ \bibinfo {pages} {1303 } (\bibinfo {year} {1980})}\BibitemShut
  {NoStop}%
\end{thebibliography}%

\end{document}